\documentclass[]{elsarticle}

\usepackage{tabularx}
\usepackage{xcolor}
\usepackage{subfigure}

\usepackage{url} 

\newcommand{\Fig}[1]{Figure~\ref{fig:#1}}
\newcommand{\Sec}[1]{Section~\ref{sec:#1}}
\newcommand{\Tab}[1]{Table~\ref{tab:#1}}

\begin{document}

\title{
The Impact of Vehicular Traffic Demand \\
on 5G Caching Architectures: a Data-Driven Study
}

\author{Francesco Malandrino$^\dagger$, Carla-Fabiana Chiasserini$^\dagger$$^\star$, Scott Kirkpatrick$^\ddag$\\
$\dagger$ Politecnico di Torino, Italy\\
$\star$ CNR-IEIIT, Italy\\
$\ddag$ The Hebrew University of Jerusalem, Israel
}

\begin{abstract}

The emergence of in-vehicle entertainment systems and self-driving vehicles, and the latters' need for high-resolution, up-to-date maps, will bring a further increase in the amount of data vehicles consume. Considering how difficult Wi-Fi offloading in vehicular environments is, the bulk of this additional load will be served by cellular networks. Cellular networks, in turn, will resort to caching at the network edge in order to reduce the strain on their core network -- an approach also known as {\em mobile edge computing}, or ``fog computing''.

In this work, we exploit a real-world, large-scale trace coming from the users of the We-Fi app in order to (i) understand how significant the contribution of vehicular users is to the global traffic demand; (ii) compare the performance of different caching architectures; and (iii) studying how such a performance is influenced by recommendation systems and content locality.
We express the price of ``fog computing'' through a metric called {\em price-of-fog}, accounting for the extra caches to deploy compared to a traditional, centralized approach.
We find that ``fog computing'' allows a very significant reduction of the load on the core network, and the price thereof is low in all cases and becomes negligible if content demand is location specific.
We can therefore conclude that vehicular networks make an excellent case for the transition to mobile-edge caching: thanks to the peculiar features of vehicular demand, we can obtain all the benefits of ``fog computing'', including a reduction of the load on the core network -- reducing the disadvantages to a minimum.
\end{abstract}

\maketitle
\thispagestyle{plain}
\pagestyle{plain}

\section{Introduction}

Back in 2010, the traffic demand of newly-introduced iPhones briefly disrupted some cellular networks~\cite{arnaud}. It is uncertain whether such disruptions are likely to happen again; however, there is no doubt that {\em if} they do happen, vehicular users will be among the main culprits.

The reason for this trend is multifold. First, vehicles carry people, and people carry multiple, data-hungry mobile devices. Second, vehicles themselves are increasingly often equipped with entertainment devices, which only add to the problem. Third, vehicles download navigation data, e.g., map updates: while this is a minor component of the overall traffic today, it is expected to increase by orders of magnitude with the introduction of self-driving vehicles, which will need much more detailed and up-to-date maps.

To make things worse, virtually {\em all} such data demand will be served by cellular networks. Indeed, most offloading solutions target pedestrian users, because their position changes relatively slowly over time and because they are more likely to be covered by such networks as Wi-Fi.

{\em Caching} is a primary way in which cellular network operators
plan to react to this demand surge. One of the most popular solutions
is to move caches as close as possible to the users, in the context of
an approach known as {\em fog computing} (a term created by
Cisco~\cite{fog}). It is expected that doing so will increase the
cache hit ratio while reducing the service latency as well as the
traffic load on the cellular core network. On the negative side, it will require deploying multiple, smaller caches. Additional help is expected from recommendation systems, whose effect is to shape the demand concentrating it around the most popular content items. Intuitively, having fewer, popular items to serve will improve caching performance.

In this context, our paper targets three main questions.

\noindent{\bf Vehicular demand.} What is the data demand generated by today's vehicular users? Which apps and services represent the most significant contributions thereto?

\noindent{\bf Caching architectures.} Given a target hit ratio, what
is the relationship between  caching architecture and  size of the
caches we need to deploy? What is the impact of moving  caches from
core-level switches to individual base stations, on the total cache size,
on the distance data must travel within the core network, and on the load thereof?
What changes if a recommendation system is in place?

\noindent{\bf Location-specific content.} Content items consumed by
future vehicular networks are expected to strongly depend on the
location -- augmented maps for self-driving vehicles being the most
obvious example. What will be the impact of this kind of content on caching?

We answer these questions using a set of real-world, large-scale measurement data, coming from users of the WeFi app~\cite{wefi}. Due to its crowd-sourced nature, our dataset includes data for:
(i) multiple apps, including video (e.g., YouTube) and maps;
(ii) multiple types of users, from pedestrian to vehicular ones;
(iii) multiple network technologies, including 3G, LTE, and Wi-Fi;
(iv) multiple operators.

We describe our dataset, as well as the additional processing we need
to perform in order to enhance the information it provides, in \Sec{dataset}. Then, in \Sec{caching} we explain how we model caching and caching architectures in our vehicular scenario. \Sec{results} presents numerical results and some relevant insights we obtain from them. Finally, \Sec{conclusion} concludes the paper and sketches future work directions.

\section{Input data}
\label{sec:dataset}

We describe the WeFi dataset we have access to in \Sec{wefi}. Then in \Sec{process} we detail the processing steps we need, in order to extract further information that is not directly included therein. Finally, \Sec{complement} explains how we complement the available information using other datasets and well-known information.

\subsection{The WeFi dataset}
\label{sec:wefi}

Our data comes from the users of an app called WeFi~\cite{wefi}.  WeFi
provides its users with information on the safest and fastest Wi-Fi
access points available at the user's location. At the same time, it collects information about the user's
location, connectivity and activity. 
WeFi  reports over seven million downloads of the app globally, and over three billion daily records. In this work, we use the dataset relative to the city of Los Angeles --  a vehicle-dominated environment. Its main features are summarized in \Tab{dataset}.

\begin{table}[]
\caption{
The Los Angeles dataset
\label{tab:dataset}
} 
\centering
\begin{tabularx}{.8\columnwidth}{|X|X|}
\hline
Metric & Value \\
\hline\hline
Time of collection & Oct. 2015\\
\hline
Total traffic & 35 TByte\\
\hline
Number of records & 81 million\\
\hline
Unique users & 64,386\\
\hline
Unique cell IDs & 47,928\\
\hline
Mobile operators & AT\&T (16,992)\\
(number of cells) & Sprint (2,764)\\
& T-Mobile (24,290)\\
& Verizon (3,882)\\
\hline
\end{tabularx}
\end{table}

\begin{figure*}[h!]
\centering
\subfigure[\label{fig:distance-cdf}]{
\includegraphics[width=.3\textwidth]{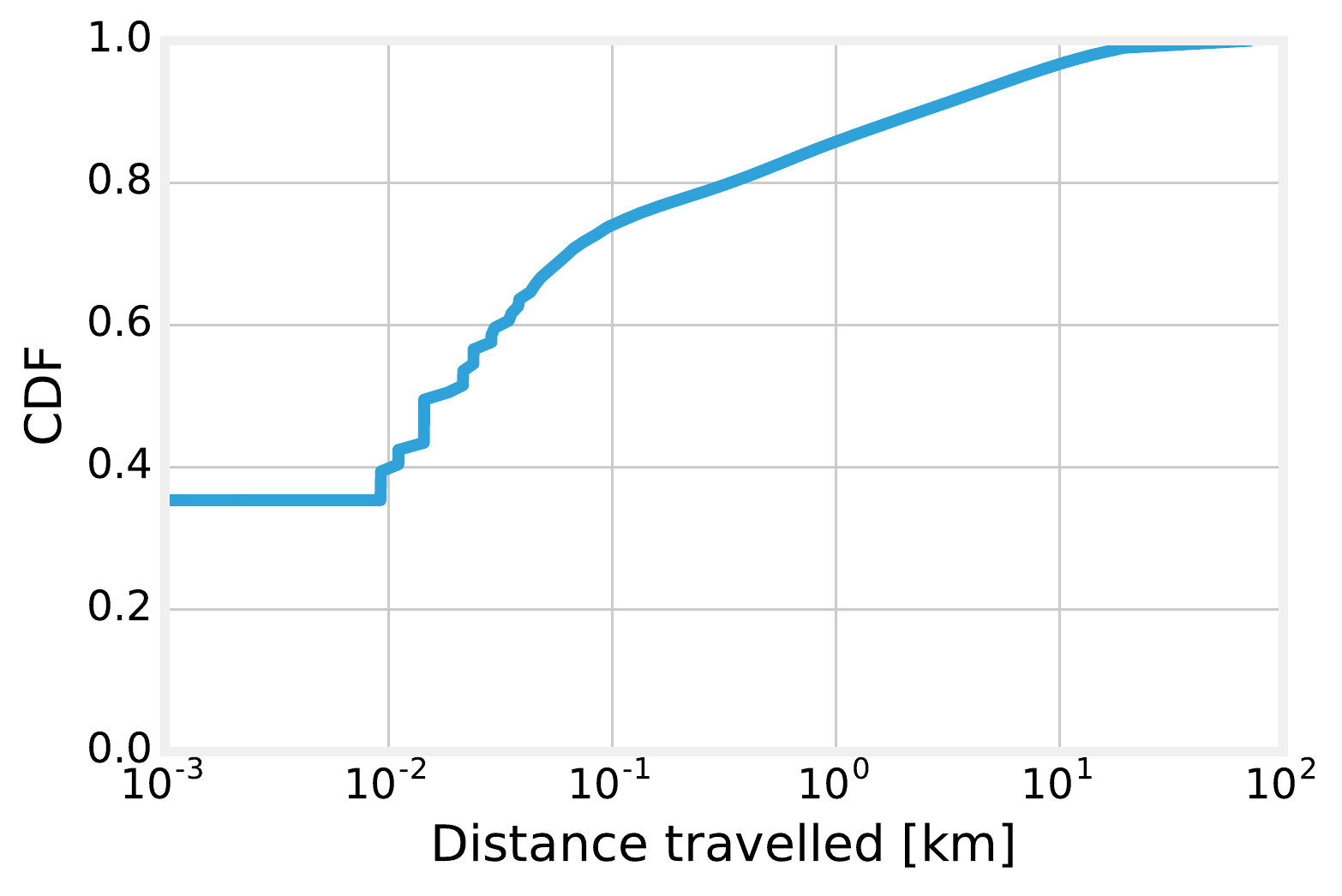}
} 
\subfigure[\label{fig:frac-veh}]{
\includegraphics[width=.3\textwidth]{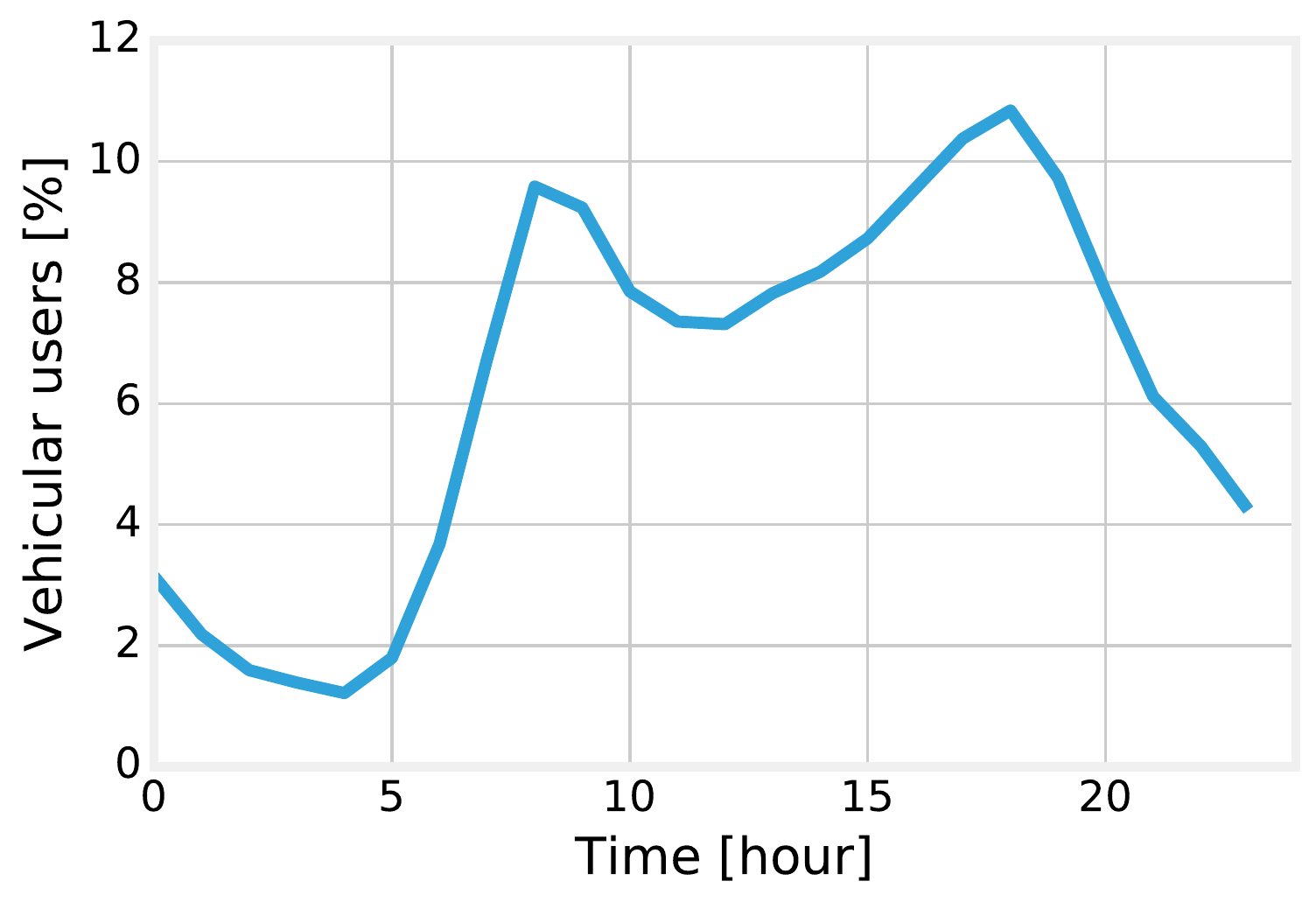}
} 
\subfigure[\label{fig:cat-pie}]{
\includegraphics[width=.3\textwidth]{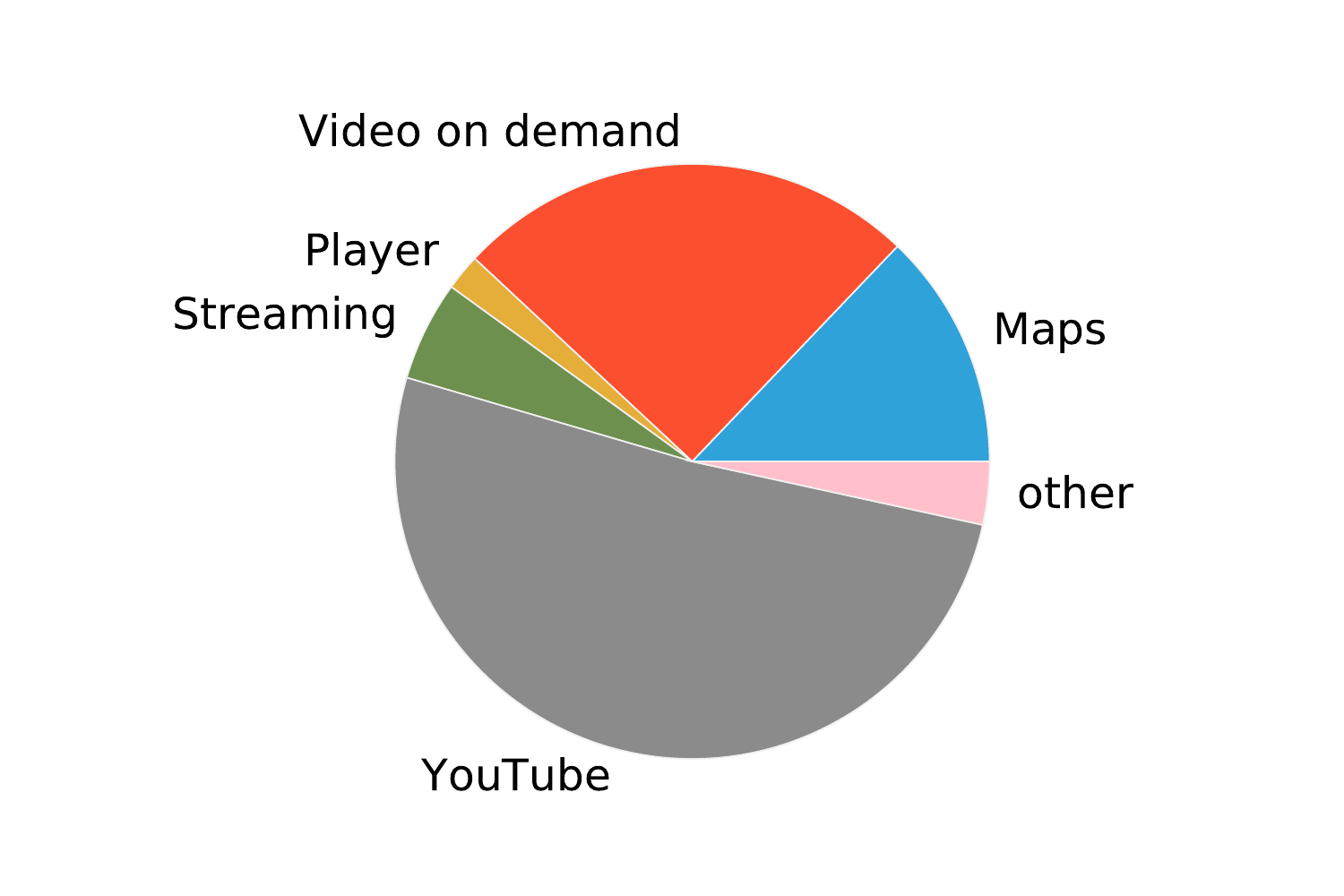}
} 
\caption{Distribution of the distance covered by users in the dataset (a); fraction of vehicular users as a function of time (b); share of data volume for the most popular app categories used by vehicular users (c).
} 
\end{figure*}

Each record contains the following information:
\begin{itemize}
\item day, hour (a coarse-grained timestamp);
\item anonymized user identifier and GPS position;
\item network operator, cell ID, cell technology and local area (LAC) the user is connected to (if any);
\item Wi-Fi network (SSID) and access point (BSSID) the user is connected to (if any);
\item active app and amount of downloaded/uploaded data.
\end{itemize}
If the location of the user or the networks she is connected to change within a one-hour period, multiple records are generated. Similarly, one record is generated for each app that is active during the same period.
The fact that location changes trigger the creation of multiple records allows us to assess whether, and how much, each user moves during each one-hour period. As we will see in \Sec{process}, this is instrumental in distinguishing between static and vehicular users. Combining this knowledge with network technology information allows us to ascertain which types of traffic cellular networks ought to worry about.

\begin{figure*}[]
\centering
\includegraphics[width=.23\textwidth]{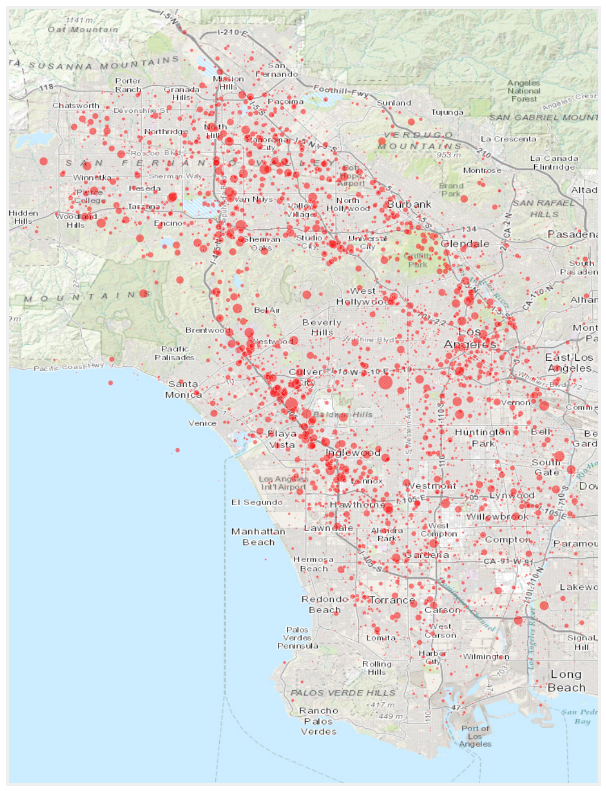}
\includegraphics[width=.23\textwidth]{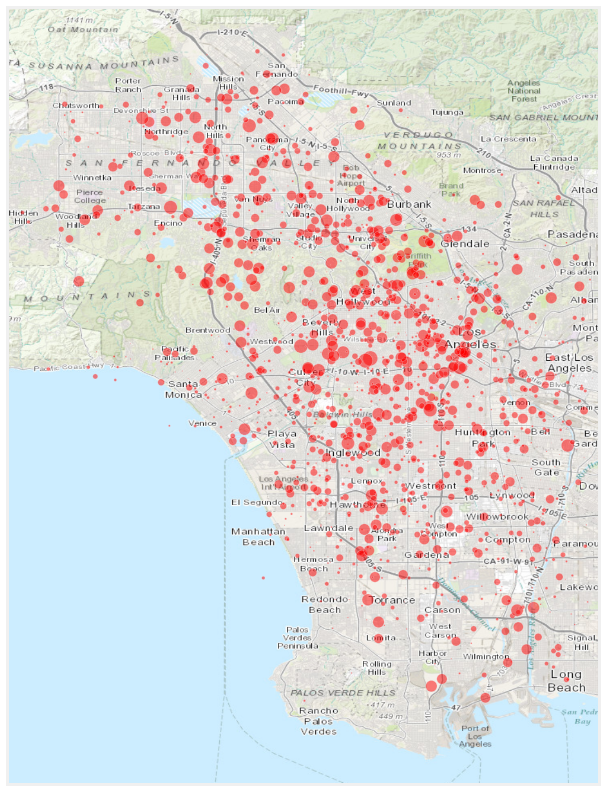}
\includegraphics[width=.23\textwidth]{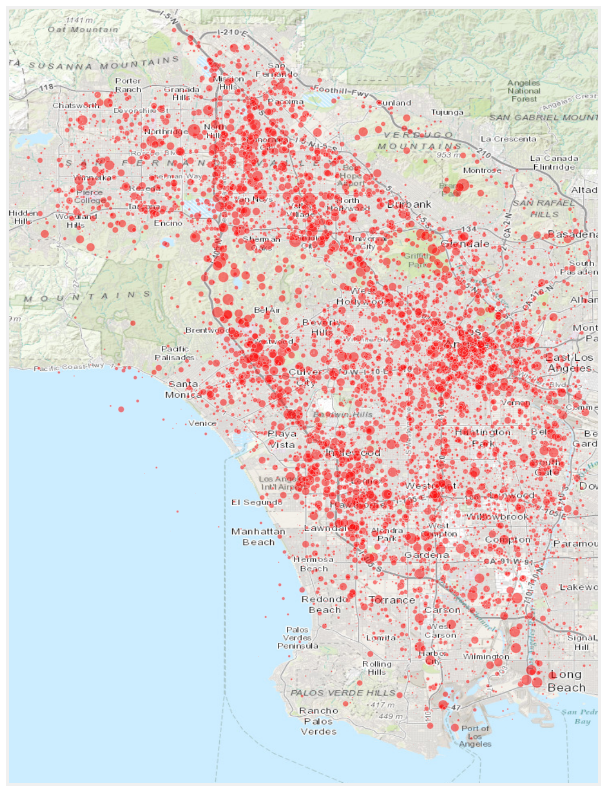}
\includegraphics[width=.23\textwidth]{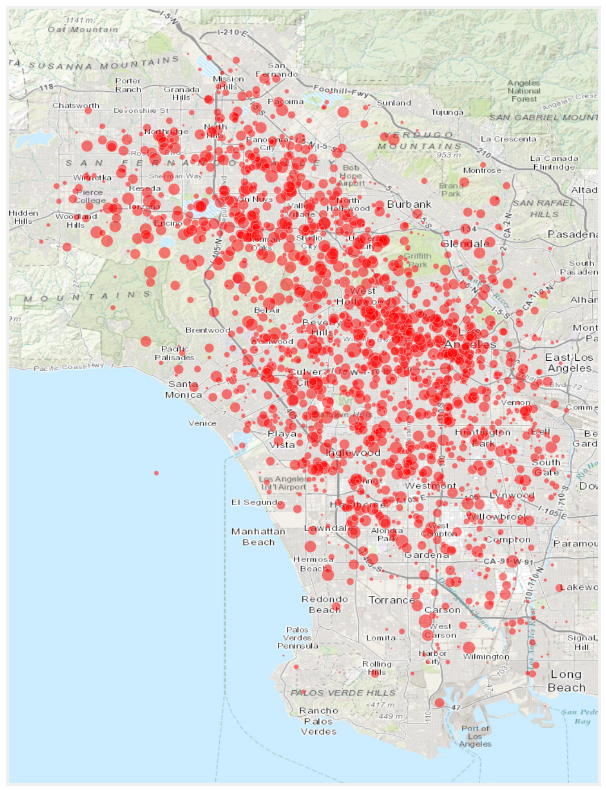}
\caption{
Left to right: deployment for AT\&T, Sprint, T-Mobile, Verizon. Each dot represents a cell, and the size of dots is proportional to the coverage area thereof, as estimated from the location of users reporting the same cell ID.
\label{fig:deployment}
} 
\end{figure*}

\Fig{deployment} shows the cell deployment of the four main operators present in our trace. We can see that all operators cover the whole geographical area we consider, but using radically different strategies. T-Mobile and, to a lesser extent, AT\&T, deploy a large number of cells, each covering a comparatively small area. Sprint and, especially, Verizon, follow the opposite approach: their networks are composed of relatively few cells, each covering a fairly large area.

This fundamental difference reflects on the topologies of each operator's core network, and potentially on the effectiveness of different caching architectures. It is worth to stress that using a real-world, crowd-sourced trace such as ours, we are able to properly account for these factors, which are typically neglected by more abstract models.

\subsection{Further data processing steps}
\label{sec:process}

From the WeFi dataset we easily identify several types of users and the content they consume.

\noindent{\bf User type.}
The WeFi app can be installed on a variety of mobile devices. The users carrying them can be static (e.g., sitting in a caf\'{e}), pedestrian (e.g., walking or jogging), or vehicular. We discriminate among these cases by looking at the {\em distance} covered by each user during each one-hour period. \Fig{distance-cdf} shows the distribution thereof: we have almost 40\% of static users, which do not move at all, a large number of pedestrian users covering moderate distance, and some users covering larger ones.

{\color{blue}
Given our focus on vehicular applications, it is important not to include pedestrian or static users in our analysis. To this end,
} 
conservatively we label as vehicular those users that travel more than 5~km in any one-hour period\footnote{Notice that the same user can be vehicular in some time periods and static in others.}. \Fig{frac-veh} shows the fraction of vehicular users as a function of time, and exhibits the familiar morning and afternoon peaks.

\begin{table}[]
\caption{
Content categories
\label{tab:categories}
} 
\centering\footnotesize
\begin{tabularx}{1\columnwidth}{|c|X|}
\hline
Category & Description \\
\hline\hline
YouTube & All class names pertaining to YouTube\\
\hline
OnDemand & On-demand video services such as Netflix, Time Warner, and ShowTime\\
\hline
RealTime & Real-time streaming, e.g., Periscope and DirectTV\\
\hline
Players & Player apps such as VLC and HTC Video\\
\hline
Weather & Most notably Weather.com\\
\hline
Maps & Most notably Google Maps\\
\hline
News & Including CNN and NBC\\
\hline
Sports & NFL, Fox Sports and the like\\
\hline
\end{tabularx}
\end{table}

\noindent{\bf Content type.}
As recalled in \Sec{dataset}, records contain an \path{app} field,
containing the class name of the active application, e.g.,
\path{COM.GOOGLE.ANDROID.APPS.YOUTUBE.KIDS}. However, we cannot use
this information directly for two main reasons. First and foremost,
different class names may correspond to the same app, e.g., both
\path{COM.GOOGLE.ANDROID.APPS.YOUTUBE.KIDS} and
\path{COM.GOOGLE.ANDROID.YOUTUBE} correspond to YouTube. Furthermore,
we are not only interested in individual apps, but also in the {\em category} they belong to, as summarized in \Tab{categories}.

It is important to point out that different content categories lend themselves to caching to radically different extents. Caches are virtually useless for real-time streaming content (while LTE broadcasting~\cite{noi-broadcasting} represents a more promising alternative). On-demand video content can be successfully cached, especially if popular. Sport and news content is even easier to cache, as there is a limited number of items that are likely to be requested (e.g., the highlights of yesterday's games). Finally, weather and map content is highly local, as users are very likely to need information about their current location.

The relative importance of the aforementioned categories is summarized in \Fig{cat-pie}. YouTube and other on-demand content dominate the vehicular traffic, while real-time streaming represents much of the rest. This is good news from the caching viewpoint, since much of the vehicular traffic is represented by content that can be successfully cached.

\begin{figure}[b!]
\centering
\includegraphics[width=0.49\textwidth]{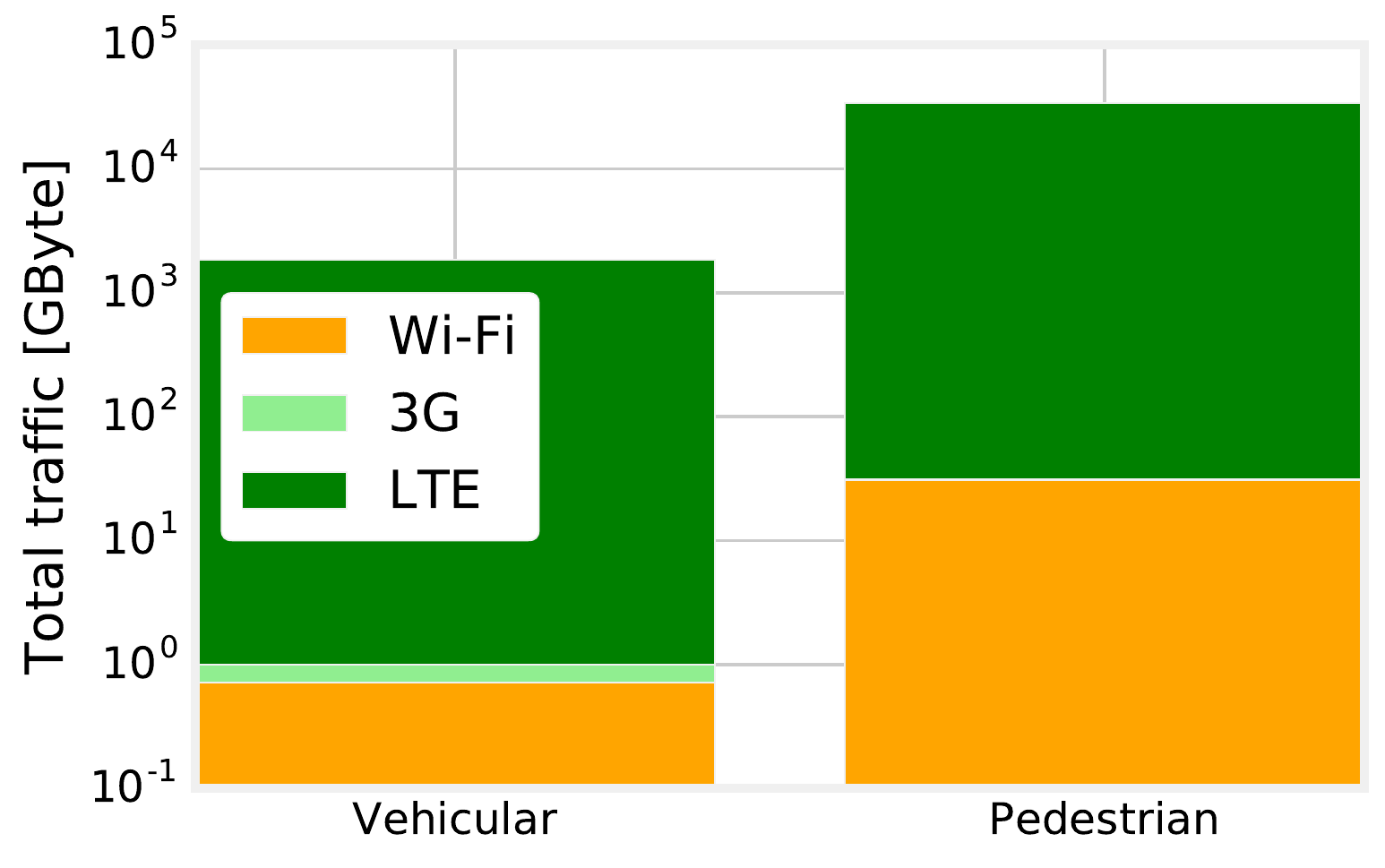}
\caption{
{\color{blue}
Breakdown of the traffic in the WeFi trace, according to the type of user originating it (vehicular or static) and the technology serving it (3G, LTE, Wi-Fi).
} 
\label{fig:mosaic}
} 
\end{figure}

{\color{blue}
\Fig{mosaic} provides further motivation to our study, allowing us to observe two important issues. First, the bulk of the traffic we observe from our trace is generated by static users (right bar), with vehicular users accounting for roughly 5\% of the total (left bar). This is going to change in the future, as in-vehicle entertainment systems and, more importantly, self-driving vehicles become more common.

The other difference between static and vehicular traffic is, instead, going to stay. The vast majority of static traffic is served through Wi-Fi (orange areas). This is not surprising, considering how we are surrounded by access points in homes, offices, and commercial activities. At the same time, the bulk of vehicular traffic is served through 3G, or (increasingly) LTE cellular networks (light- and dark-green areas respectively).
} 

In other words, if static (or pedestrian) users increase the amount of data they consume, much of this data is likely to be served by -- or offloaded to -- a Wi-Fi network; conversely, cellular networks are alone in facing the increase of data demand from vehicular users and vehicles.

\subsection{Network topology and content demand}
\label{sec:complement}

There are two types of information that are altogether missing in our WeFi dataset: network topology (both access and core), and content demand. In the following, we explain how we reconstruct this information using other existing datasets and/or common knowledge.

\begin{figure}[h!]
\centering
\includegraphics[width=0.4\textwidth]{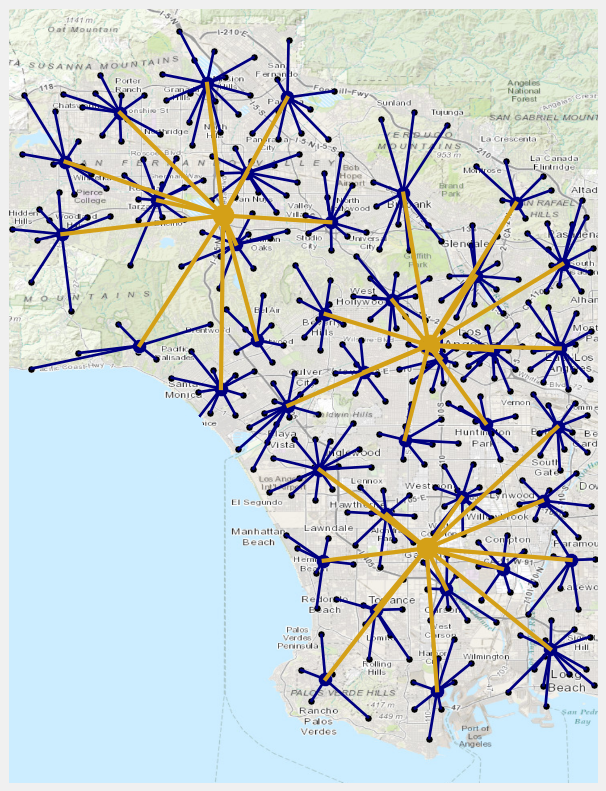}
\caption{
Assumed core network topology for Verizon. Black circles correspond to base station rings, blue ones to aggregation-layer pods, and yellow ones to core-layer switches. Lines represent links.
\label{fig:topo}
} 
\end{figure}

\noindent{\bf Network topology.}
In order to study the effectiveness of different caching architectures, we need information about how base stations are connected to each other. Sadly, such information is not only absent from the WeFi dataset, but virtually impossible to obtain for any network. Indeed, this is highly sensitive information for network operators.
We estimate the position of base stations from the users' locations:
\begin{enumerate}
\item from each record, we extract the ID of the cell the user is connected to and her latitude/longitude coordinates;
\item the convex hull of these locations corresponds to the cell coverage area (notice that such areas can and do overlap);
\item we assume base stations sit at the baricenter of each convex hull.
\end{enumerate}

As for the core network, we assume, as in~\cite{softcell}, a tree topology where:
\begin{itemize}
\item base stations are grouped into {\em rings} of ten;
\item rings are connected to aggregation-layer {\em pods};
\item pods are connected to {\em core}-level switches.
\end{itemize}
Finally, we assume completely separate network topologies for each operator.

\Fig{topo} shows the core network topology we assume for Verizon, which
{\color{blue}
has roughly 3,882 cells (as per \Tab{dataset}), $\lceil\frac{3882}{10}\rceil=389$~base station rings, $\lceil\frac{389}{10}\rceil=39$~aggregation-layer pods and $\lceil\frac{39}{10}\rceil=4$ core switches. The topologies of the other mobile operators are generated in the same way.
} 

\noindent{\bf Per-content item demand.}
Our dataset tells us how many users use, e.g., YouTube, and how much data they consume. However, it contains no information about {\em which} of the countless YouTube videos they are watching, which is crucial to study the effectiveness of caching schemes.
We cope with this limitation through different approaches, depending on the content category:
\begin{itemize}
\item RealTime, Players: each request refers to a different content ID, modeling the fact that no caching is possible;
\item YouTube, OnDemand: the content ID is extracted from the YouTube
  measurement~\cite{youtube}, with a probability that is proportional
  to the number of each video's views;
\item News, Sports: with probability 0.9, the content item is selected
  from 50 popular ones, otherwise, a new content ID is generated;
\item Meteo, Maps: with probability 0.9 the item is selected from 10
  location-specific ones, otherwise, a new content ID is generated.
{\color{blue}
The definition of News/Sports and Meteo/Maps content items reflects the fact that most of the times, the requested item comes from a fairly restricted pool of popular items, e.g., current breaking news or yesterday's game highlight; however, users do occasionally need other items (e.g., older news) that are worthless to cache. Additionally, for Meteo and Maps, the popular items will be location-specific, e.g., a map of the local area or weather forecasts for the current location.
} 

\end{itemize}
The above assignment policy reproduces the qualitative differences
between content categories, and therefore the different ways each
lends itself to caching. Also, note that content items belonging to
different applications are always considered to be different items,
{\color{blue}
whose size comes from the real-world trace, which includes information on the amount of downloaded data.
} 

\begin{figure}[]
\centering
\includegraphics[width=.8\columnwidth]{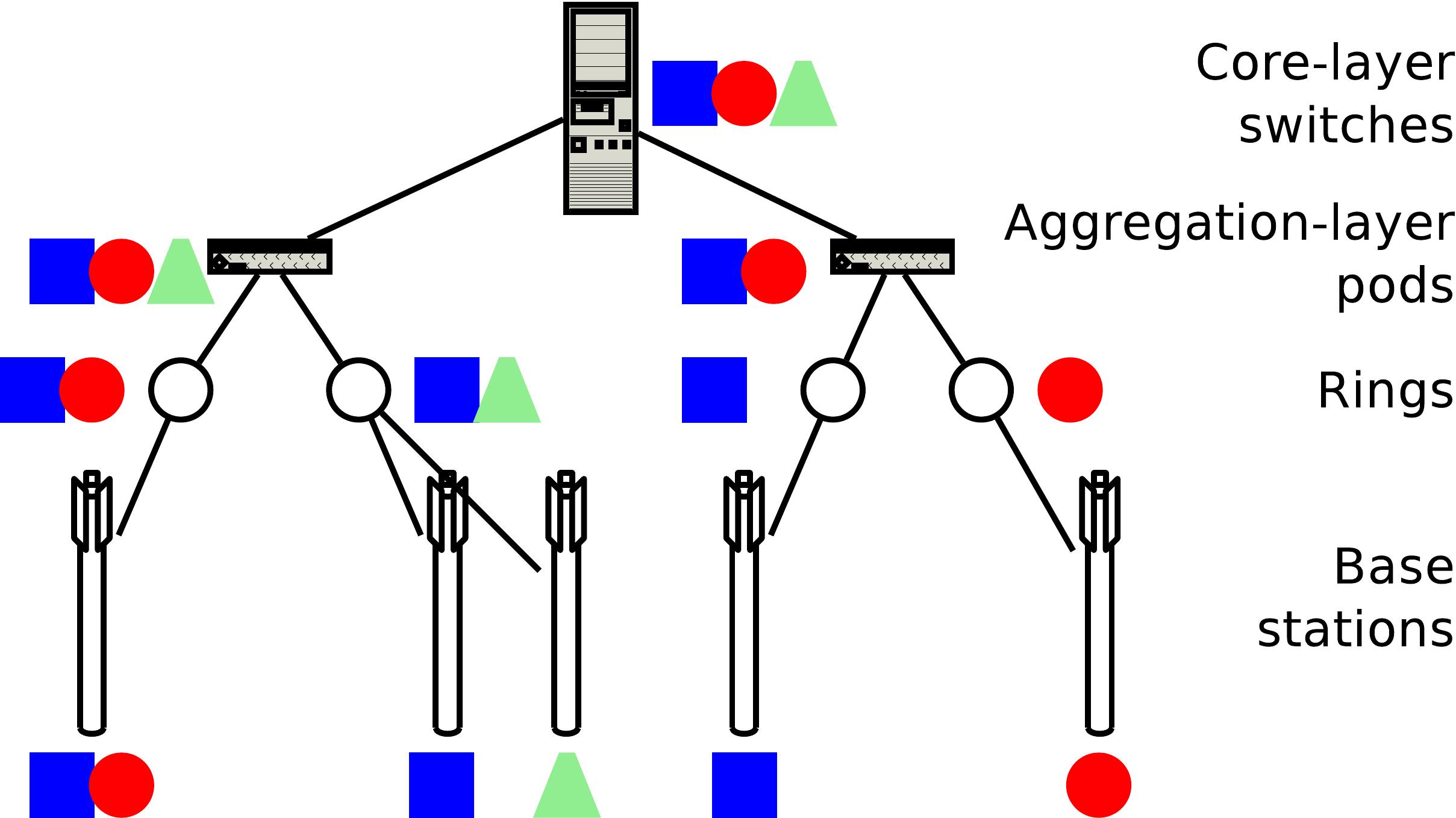}
\caption{In this simplified network architecture, base stations are
  connected to rings, rings to aggregation pods, and the latter to a
  core switch. Shapes correspond to cache-worthy content items. The
  total cache size is 6 if caches are deployed at the base
  stations or at the rings; it decreases to 5 if caches are moved to
  aggregation pods, 
and to 3 if they are located at the core switch.\label{fig:archi}
} 
\end{figure}

\section{Caching architectures}
\label{sec:caching}

Our purpose is
{\color{blue} not
 to evaluate  caching {\em policies},
} 
i.e., how to choose the content to cache, rather to evaluate cache {\em architectures}, i.e., at which level of the network topology caches should be deployed.
Four options are possible:
\begin{itemize}
\item individual {\em base stations}: each base station has its own cache, bringing the fog-computing vision to its extreme;
\item base station {\em rings}: caches are shared among the base stations (typically around ten) connected by the same ring, reducing the number of caches to deploy;
\item {\em aggregation-layer pods}: they typically serve hundreds of base stations within a fairly wide area; this represents a more centralized caching architecture;
\item {\em core-layer switches}: the most centralized caching architecture.
\end{itemize}

Given the user demand information, we consider a {\em target} hit ratio, and seek to determine the cache size needed to achieve such a ratio under different architectures.
More precisely, we proceed as follows:
\begin{enumerate}
\item we keep track of the popularity (i.e., number of requests) of each content item within each cell;
\item we sort the item/cell pairs by decreasing popularity;
\item we mark as {\em cache-worthy} enough pairs to guarantee the target hit ratio, starting from the most popular ones;
\item we identify the location at which cache-worthy content items should be stored;
\item we add at most one copy of the cache-worthy content item at said location;
\item we evaluate the total cache size needed.
\end{enumerate}
The network node at which content copies are stored (as per item 4 above) depends on the current caching architecture: if caches are deployed at base stations, it is the base station itself; otherwise, it is the core network entity (ring, aggregation pod, core-layer switch) serving that base station.

\Fig{archi} exemplifies the relationship between caching architecture
and cache size. The closer caches are to base stations and end-users,
the more likely we are to cache multiple copies of the same content
item (at different locations), thus increasing the total cache
size. On the other hand,
{\color{blue}
the size of the {\em individual}
} 
caches that are closer to the end-users tend to be smaller.
Depending on the technology they use, the cost of caches can grow faster than linearly with their size. In these cases, deploying several small caches can result substantially cheaper than deploying a few, larger ones. Note that a similar problem has been studied in the context of content delivery networks (CDNs)~\cite{cdn1,cdn2}.

\subsection{Performance metrics}
\label{sec:metrics}

\noindent{\bf Price-of-fog.}
We can formally define the price-of-fog metric as the ratio of the cache size to deploy under a given architecture to the cache size to deploy at the core switches. In the example case of \Fig{archi}, the price-of-fog is~$\frac{5}{3}\approx 1.67$ when placing caches at aggregation pods, and~$\frac{6}{3}=2$ when placing them at base stations or at the rings. Clearly, content popularity distribution and content locality have a major impact on the price-of-fog.

Suppose that exactly the same set of content items were deemed cache-worthy at all base stations -- perhaps as a consequence of an effective recommendation system. In the network of \Fig{archi}, the price-of-fog would raise as high
{\color{blue}
as~$5$, with one copy of the same content deployed at each base station
} 
-- and much higher in real networks, where core nodes have more descendants. At the other extreme, if the set of cache-worthy content items at every base station were disjoint, the price-of-fog would drop to~$1$, the lowest possible value. Indeed, one of the main contributions of our paper is to assess to which of these extreme cases current {\em and} future vehicular networks are closer.

\begin{figure*}[b]
\centering
\subfigure[\label{fig:cachecdf-bs}]{
\includegraphics[width=.3\textwidth]{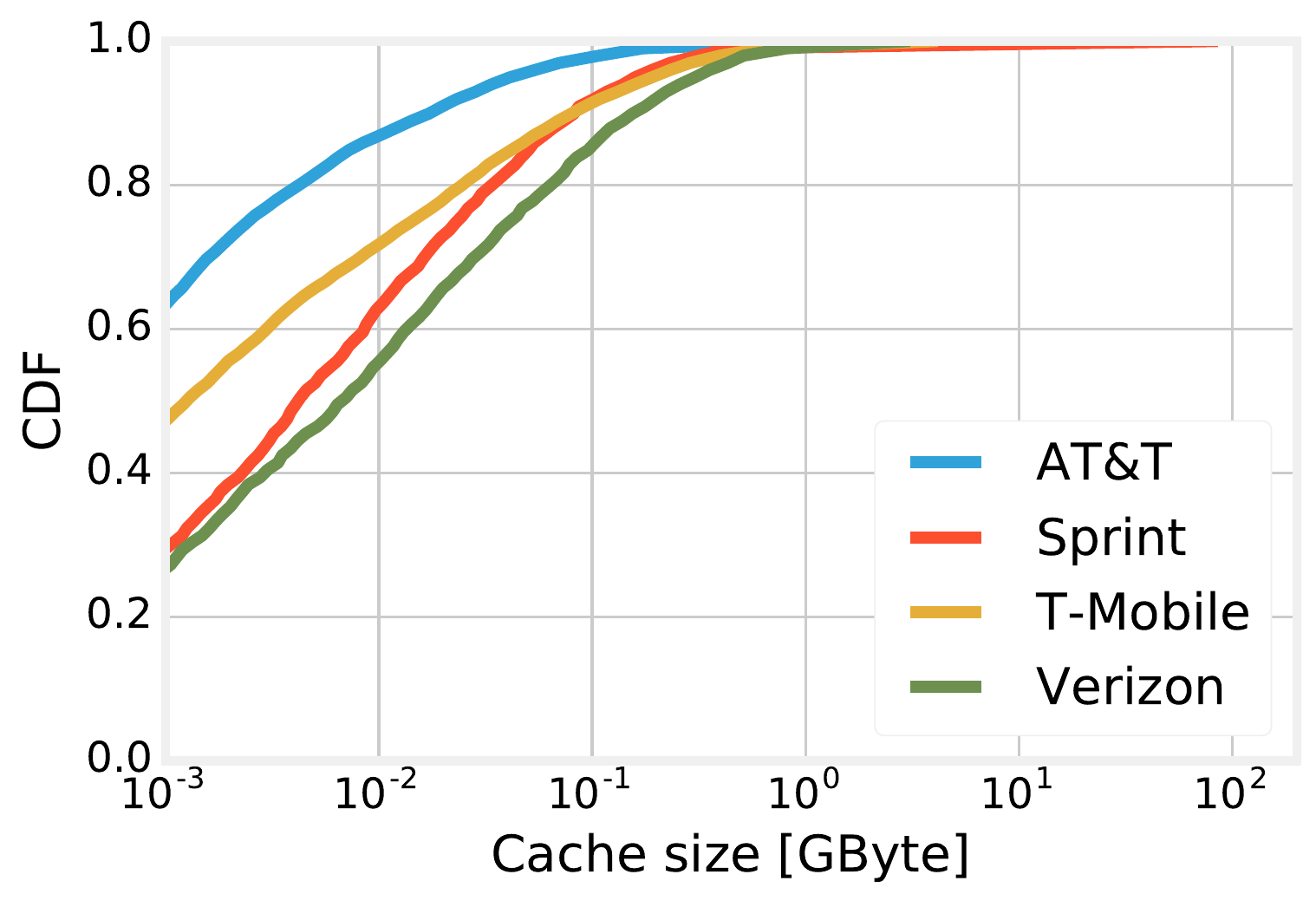}
} 
\subfigure[\label{fig:cachecdf-pod}]{
\includegraphics[width=.3\textwidth]{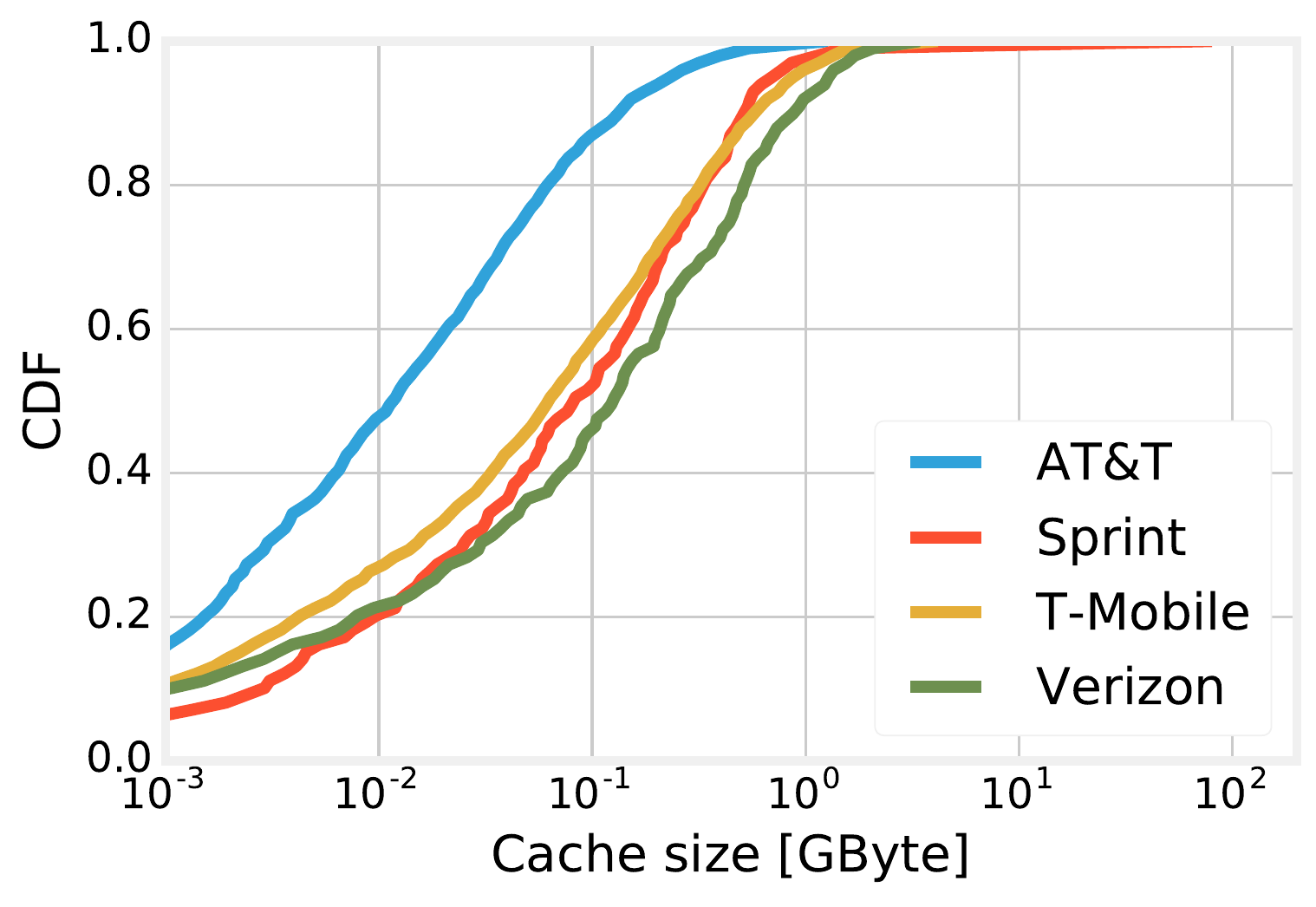}
} 
\subfigure[\label{fig:cache-totsize}]{
\includegraphics[width=.3\textwidth]{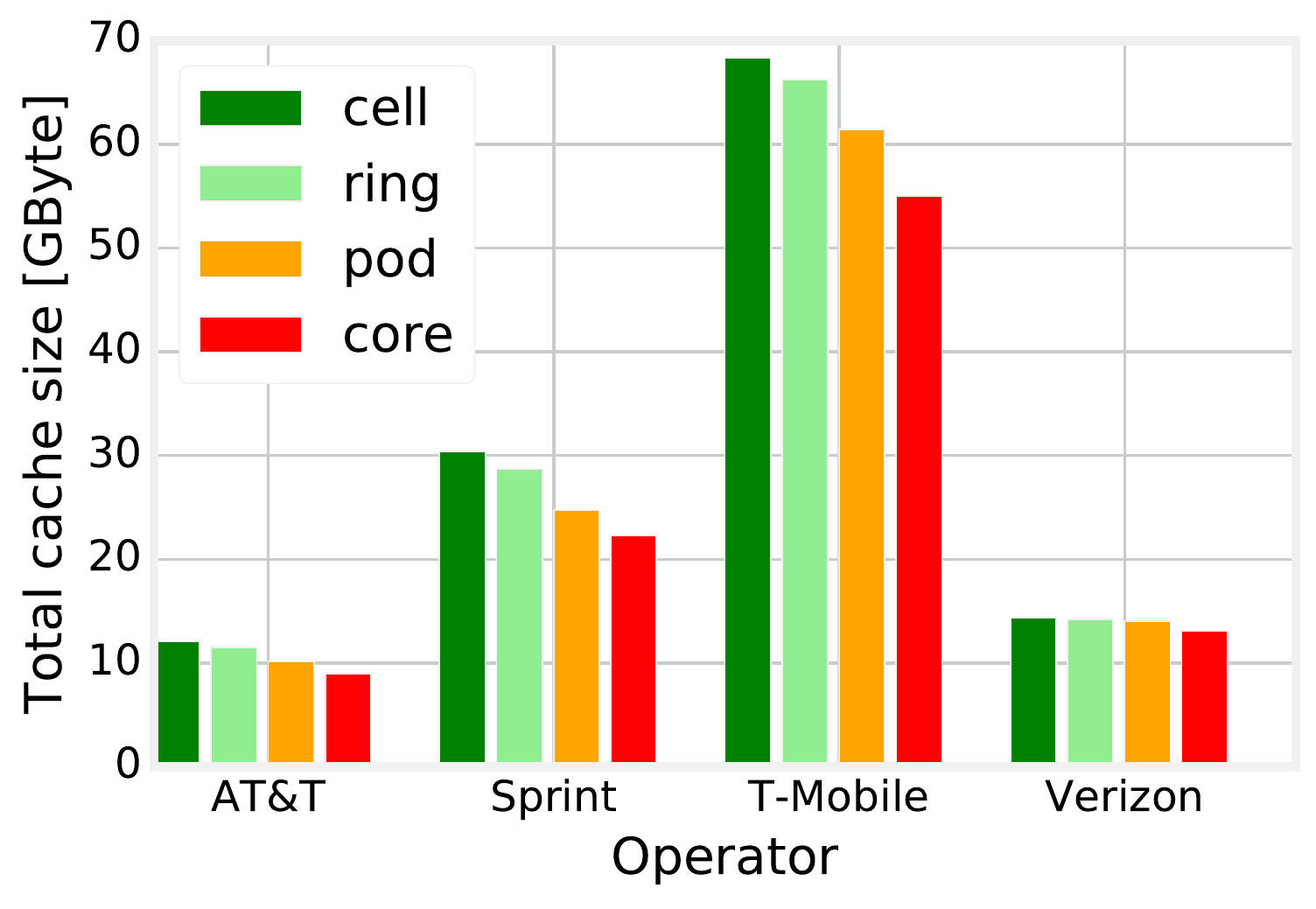}
} 
\caption{Distribution of the cache size when they are deployed at base stations (a) and rings (b); total cache size for different architectures (c).
\label{fig:cachesize}
} 
\end{figure*}

\noindent{\bf Distance travelled by data.}
Fog computing essentially means moving data closer to the users, thus reducing the load on the core network. We quantify this effect by measuring the physical distance between the network node at which content items are cached (e.g., aggregation-layer pods or core-layer switches) and the base station serving it.

\subsection{Recommendation systems and local content}
\label{sec:recloc}

We study two factors that can alter the content demand and the distribution thereof: recommendation systems and the presence of location-specific content. The latter is expected to become a dominant factor in the near future, especially for vehicular applications.

\noindent
{\em Recommendation systems} have the high-level effect of concentrating the demand towards the most popular items. To model this, we first track the top 5\% most popular content items for each app; then, for each request, we switch the requested content to one of those popular ones with a probability~$p$. The higher~$p$, the stronger the bias towards popular content.

\noindent
In the case of {\em location-specific content}, we create~$5$ new content items specific to each cell; then, for each request, we switch the requested content to one of those local ones with a probability~$q$. The higher~$q$, the stronger the correlation between user location and content demand.

\section{Numerical results}
\label{sec:results}

\begin{figure*}[]
\centering
\subfigure[\label{fig:rec-pof}]{
\includegraphics[width=.3\textwidth]{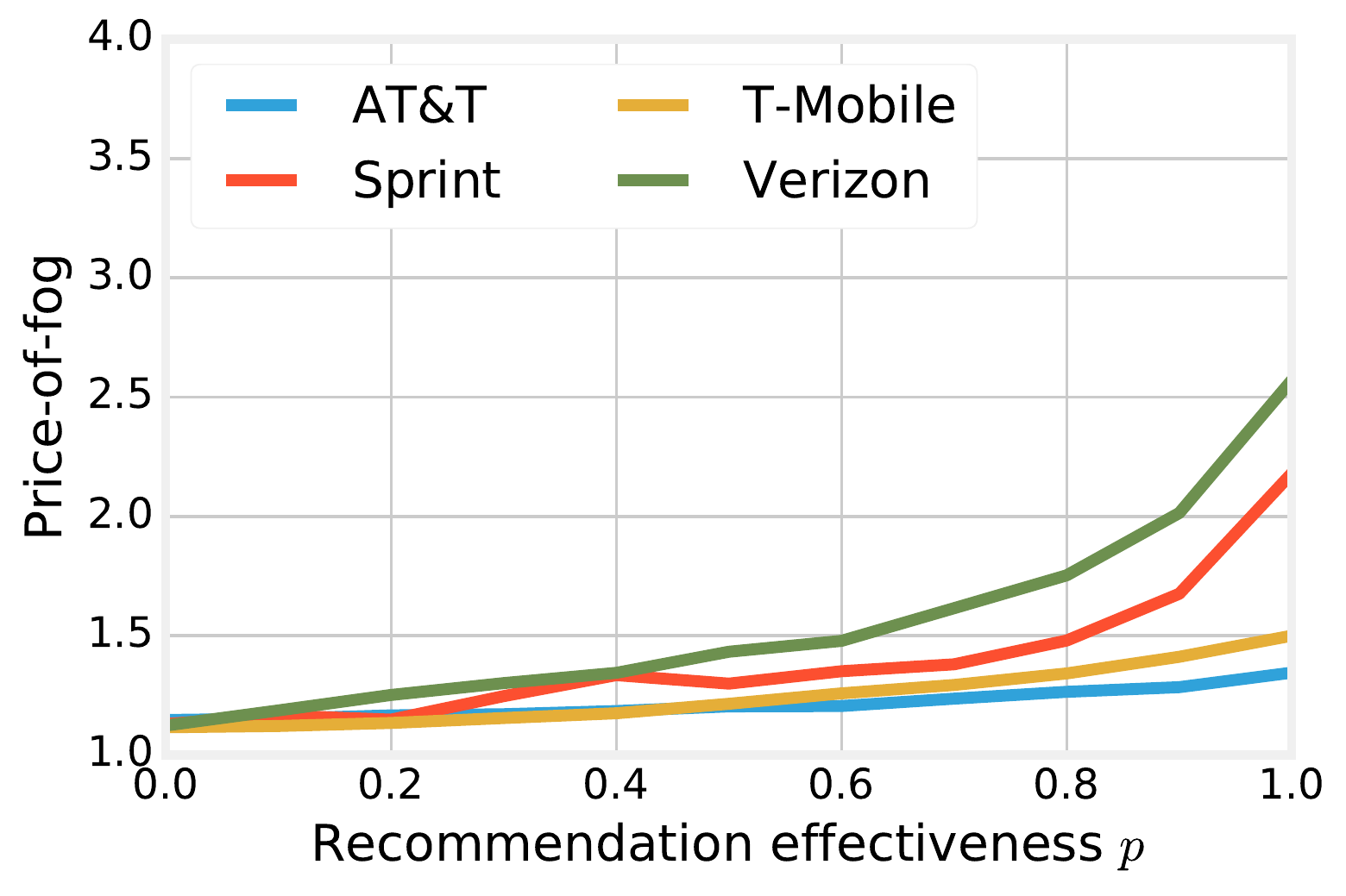}
} 
\subfigure[\label{fig:rec-cachesize}]{
\includegraphics[width=.3\textwidth]{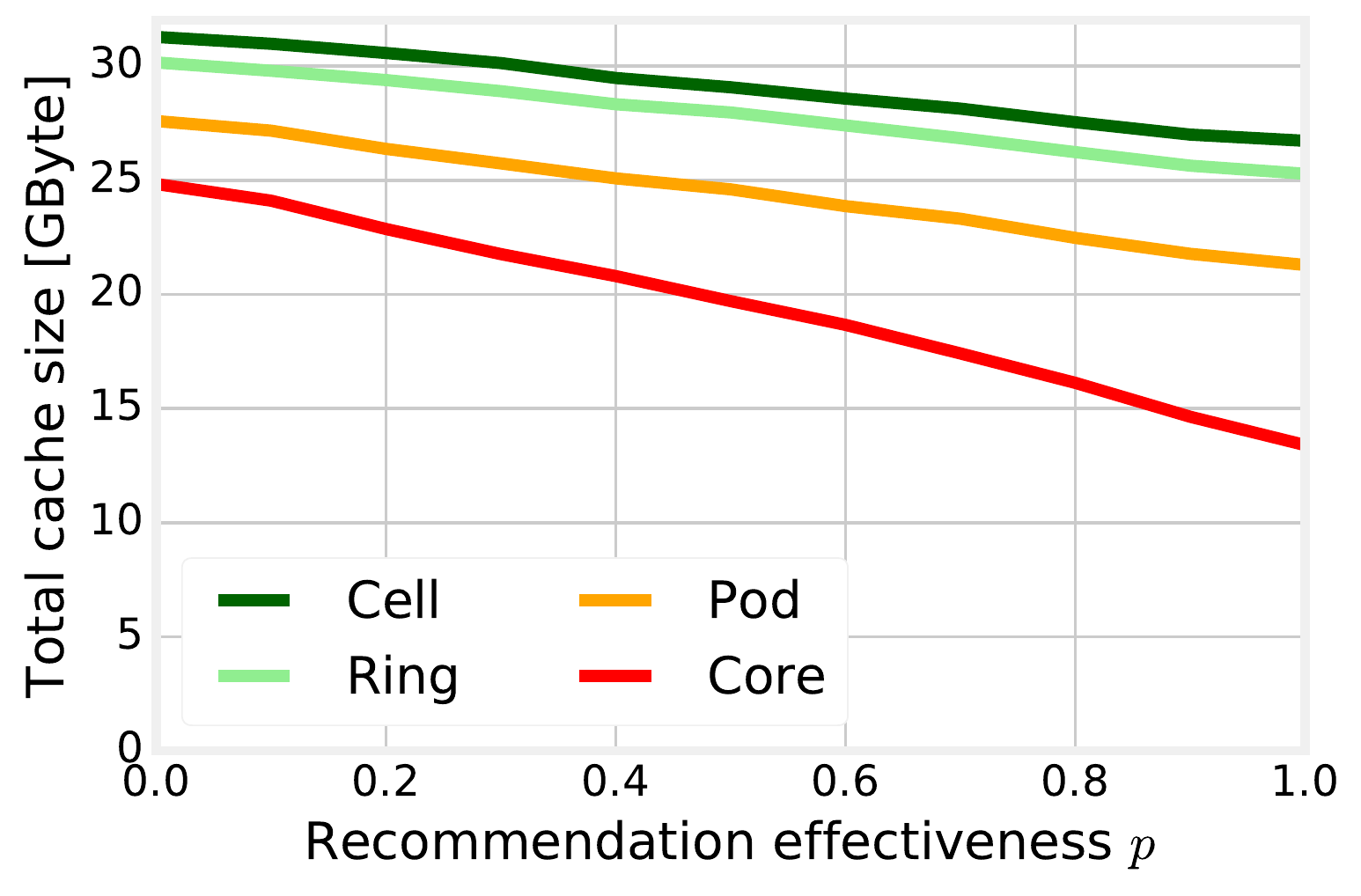}
} 
\subfigure[\label{fig:rec-bars}]{
\includegraphics[width=.3\textwidth]{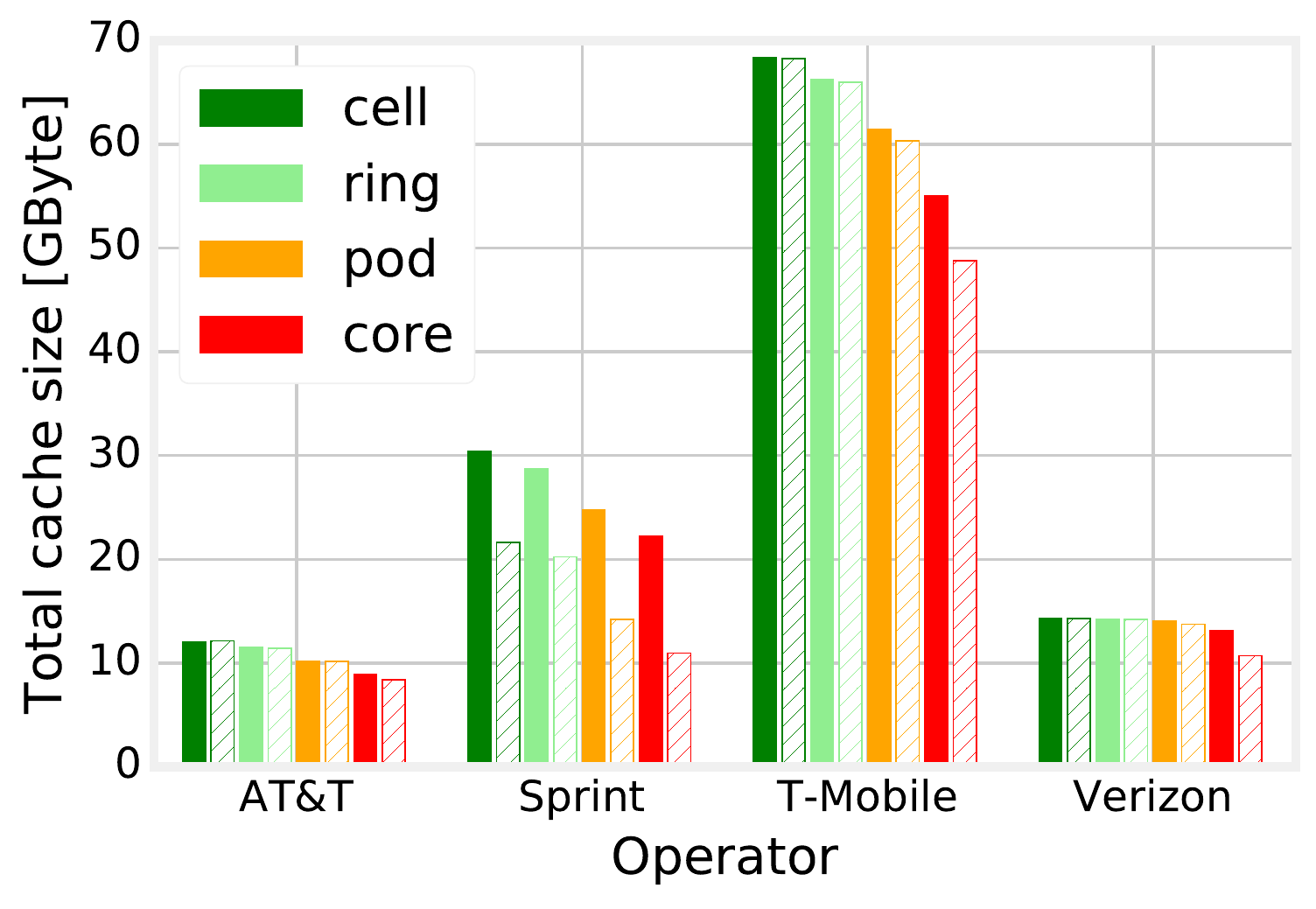}
} 
\caption{Recommendation system: price-of-fog
{\color{blue}
when caches are deployed at base stations
} 
(a); total cache size averaged
over the different operators, as a function of~$p$ (b); per-operator breakdown when~$p=0$ (solid bars) and~$p=0.5$ (bars with pattern) (c).
} 
\end{figure*}

A first aspect we are interested into is cache size. For each cache architecture, we are interested in (i) the distribution of cache sizes, and (ii) the total size thereof.
{\color{blue}
Unless otherwise specified, we keep our target hit ratio fixed to~$0.5$.
} 

Comparing the distributions in \Fig{cachecdf-bs} and
\Fig{cachecdf-pod}, we can see that the closer caches are to the end users, the smaller their size becomes -- consistently with what one might intuitively expect.
Interestingly,
there are major differences between operators: as shown in
\Fig{deployment}, Verizon has fewer cells with larger coverage areas,
therefore, it would need to deploy larger caches. T-Mobile, on the other hand, has many smaller cells, and therefore smaller caches.

Moving to the total cache size, \Fig{cache-totsize} highlights that both the total cache size {\em and} how it changes across caching architectures strongly depend on the operator and its network. T-Mobile, with its numerous small cells, has to deploy the most caches, followed by Sprint with its few bigger ones. The other operators follow intermediate approaches, and have smaller total cache sizes.

As for the price-of-fog metric defined in \Sec{metrics}, it is actually quite modest, ranging between 1.15 for Verizon and 1.25 for AT\&T. In other words, even considering the {\em current} demand of {\em current} networks, mobile operators (and their users) can reap the benefits of fog at the cost of a moderate increase in the total cache size they need to deploy.
{\color{blue}
Notice that, because the price-of-fog is defined as a ratio between cache sizes, price-of-fog values remain comparable across architectures when absolute cache sizes (shown in \Fig{cachesize}) at the different layers change by the same factor.
} 

\noindent{\bf Recommendation system.}
We now assume that there is a recommendation system in place, as described in \Sec{recloc}, and study the effect of the $p$-value modeling its effectiveness. Somehow surprisingly, the price-of-fog depicted in \Fig{rec-pof} {\em increases} as~$p$ grows
{\color{blue}
-- this effect is especially evident for Sprint and Verizon, but can be observed for all operators.
} 
In other words, an effective recommendation system makes the fog computing approach more onerous in terms of required cache size.

Recall, however, that the price-of-fog is a ratio between two size values. As we can see from \Fig{rec-cachesize}, cache size {\em decreases} as~$p$ grows, for {\em all} caching architectures. However, the size of core-level caches decreases faster, hence the growing price-of-fog.

It is also interesting to point out that, as we can see from \Fig{rec-bars}, both the decrease in cache size and the price-of-fog strongly depend on the operator and its network topology. As an example, T-Mobile reaps significant benefits when caches are deployed at the core level and negligible ones otherwise, while Sprint experiences a decrease in cache size under all architectures.
This is again due to the differences in network deployments, shown in \Fig{deployment}. Cells covering very small areas, such as in the case of T-Mobile, are unlikely to be a good location to place a cache.

\begin{figure*}[]
\centering
\subfigure[\label{fig:loc-pof}]{
\includegraphics[width=.3\textwidth]{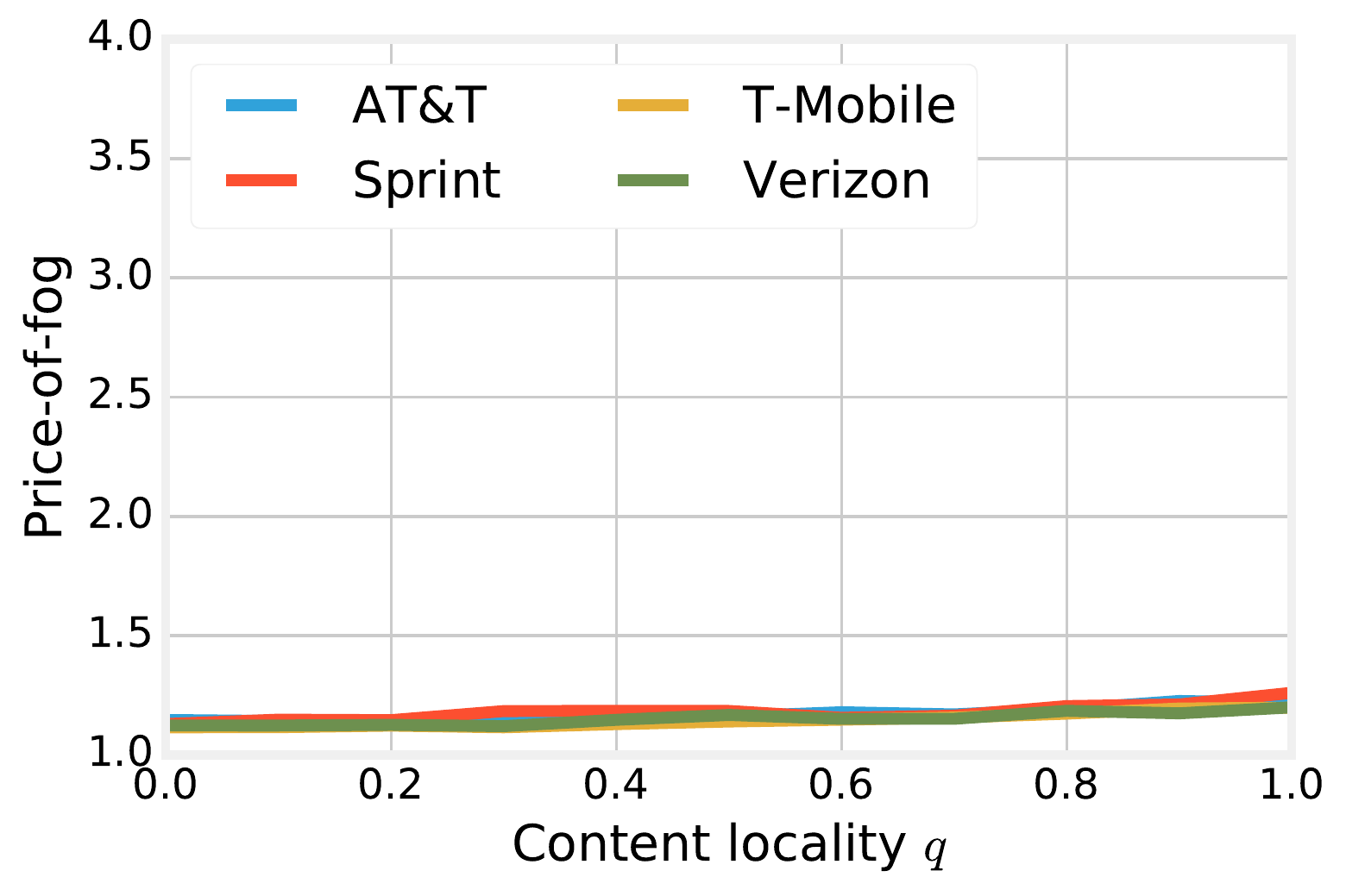}
} 
\subfigure[\label{fig:loc-cachesize}]{
\includegraphics[width=.3\textwidth]{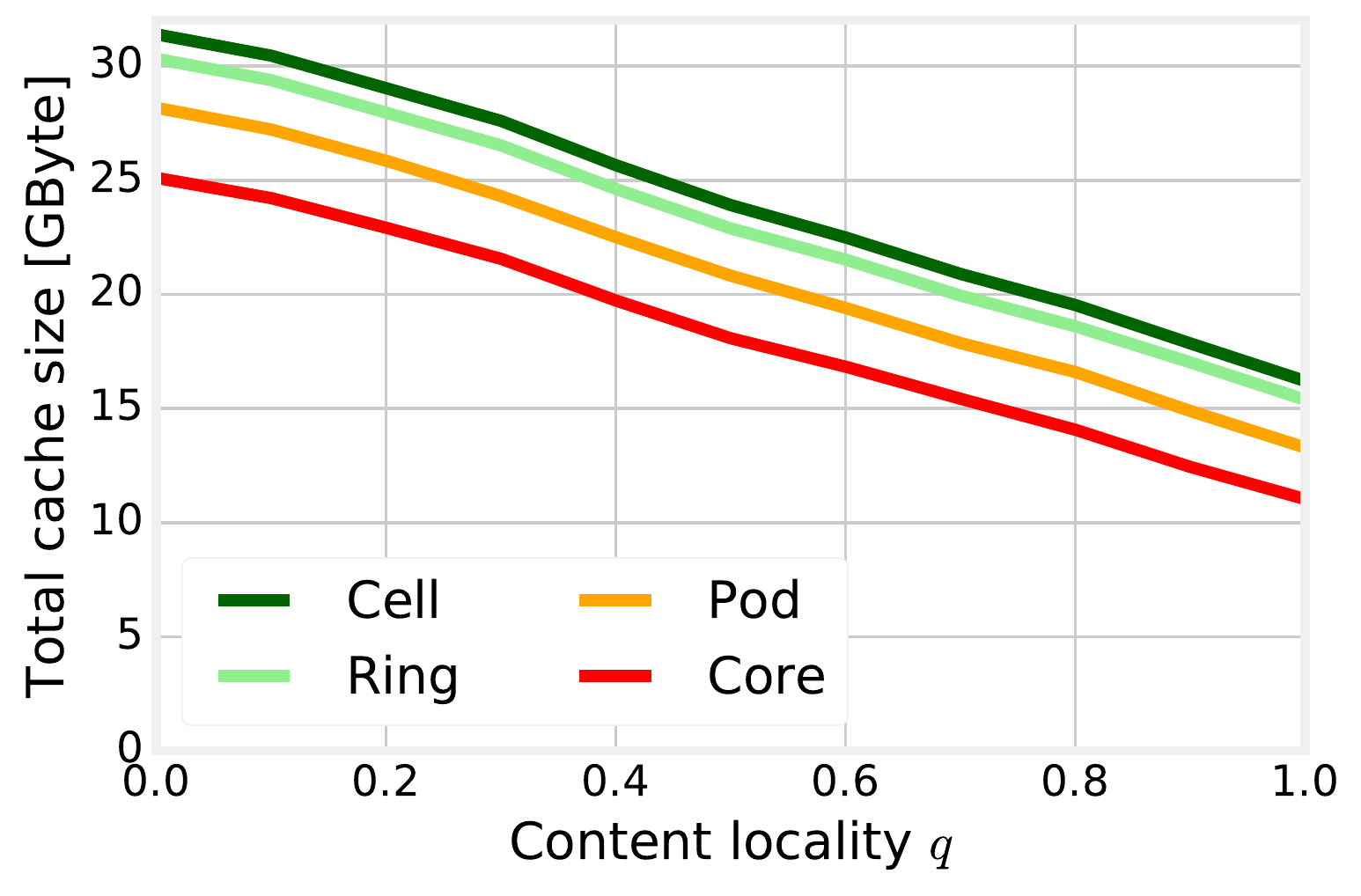}
} 
\subfigure[\label{fig:loc-bars}]{
\includegraphics[width=.3\textwidth]{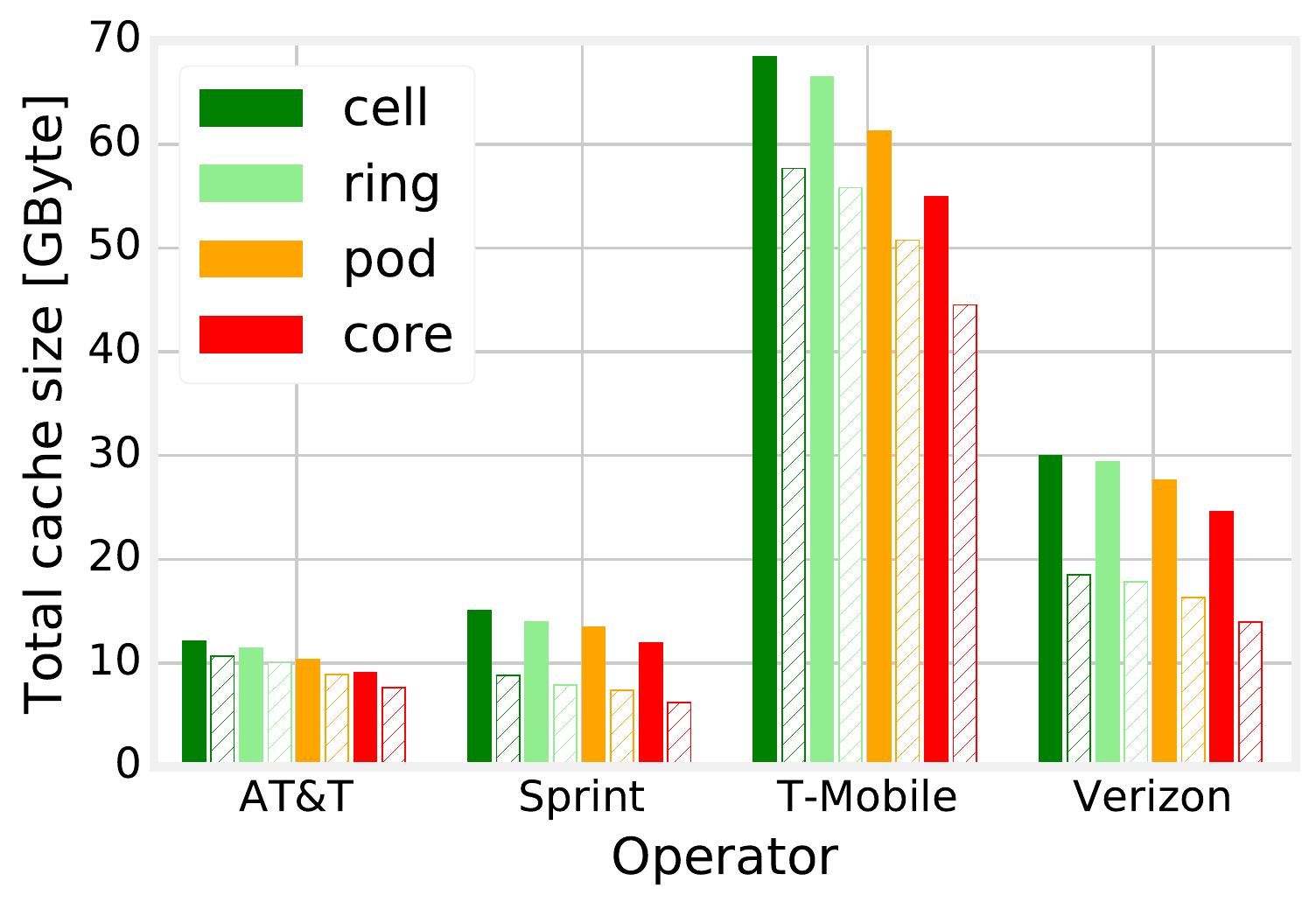}
} 
\caption{Location-specific content: price-of-fog
{\color{blue}
when caches are deployed at base stations
} 
(a); total cache size averaged
over the different operators, as a function of~$q$ (b); per-operator breakdown when~$q=0$ (solid bars) and~$q=0.5$ (bars with pattern) (c).
} 
\end{figure*}

\noindent{\bf Location-specific content.}
Recall that, as mentioned in \Sec{recloc}, the $q$-value expresses how strong the correlation between location and content demand is. Comparing \Fig{loc-pof} to \Fig{rec-pof} above, we can clearly see that the price-of-fog is (i) much lower, and (ii) virtually constant for all values of~$q$. At a high level, this tells us that if demand and location are strongly correlated, then embracing  fog computing-based caching comes at virtually no penalty,
{\color{blue}
in the real-world scenarios we consider.
} 

Consistently, \Fig{loc-cachesize} shows that cache sizes steadily decrease as~$q$ grows, for all caching architectures. Also notice, from \Fig{loc-bars}, that the effect has roughly the same magnitude for all operators.

We now consider a different performance metric, namely, the {\em distance} travelled by content items from the cache they are stored at to the base station needing them, using reference topologies such as the one described in \Fig{topo}. Clearly, a shorter distance implies less traffic on the core network and less strain on its switches and routers.

\begin{figure*}[]
\centering
\subfigure[\label{fig:dist-rec-ring}]{
\includegraphics[width=.3\textwidth]{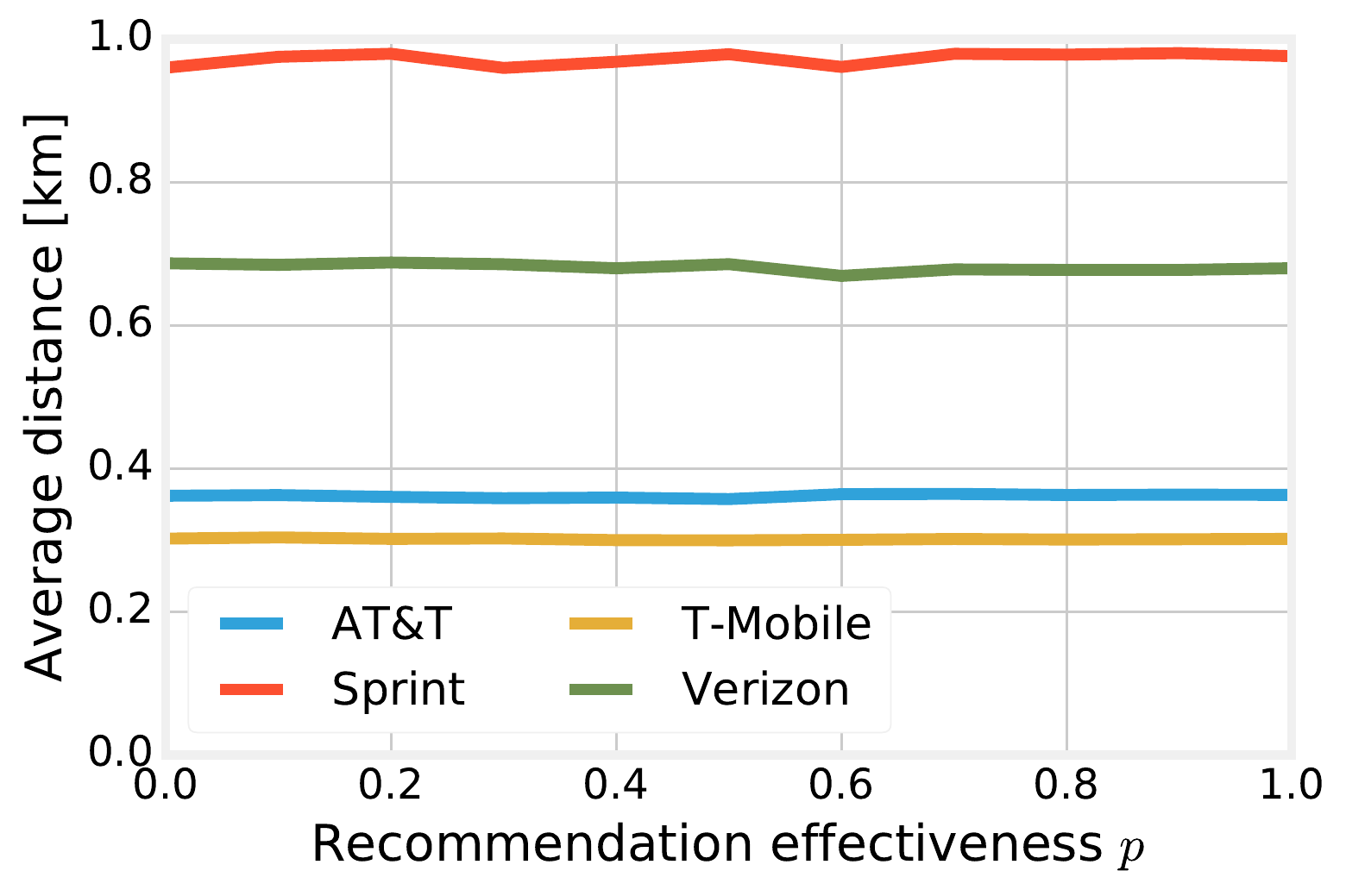}
} 
\subfigure[\label{fig:dist-rec-pod}]{
\includegraphics[width=.3\textwidth]{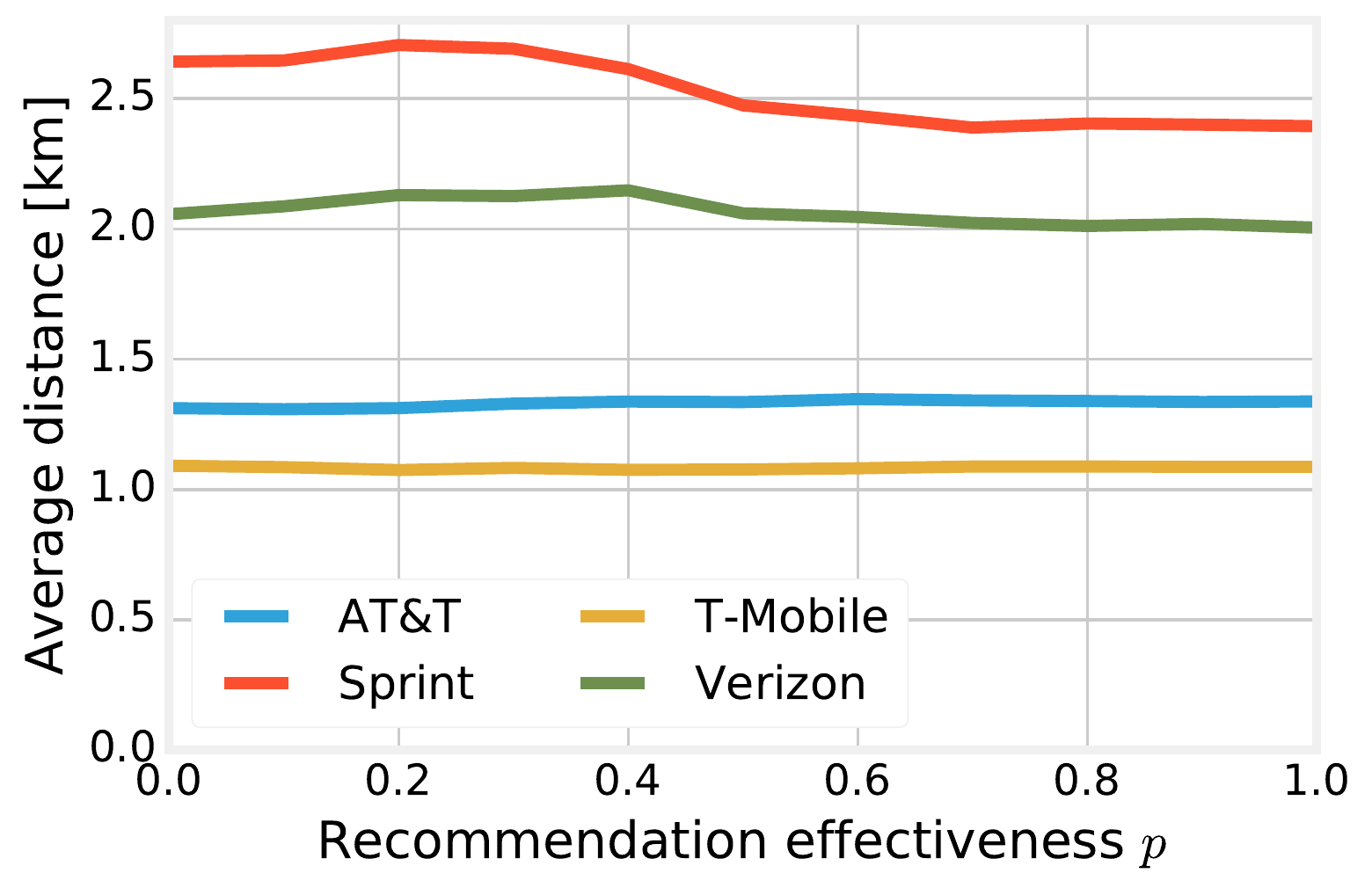}
} 
\subfigure[\label{fig:dist-rec-core}]{
\includegraphics[width=.3\textwidth]{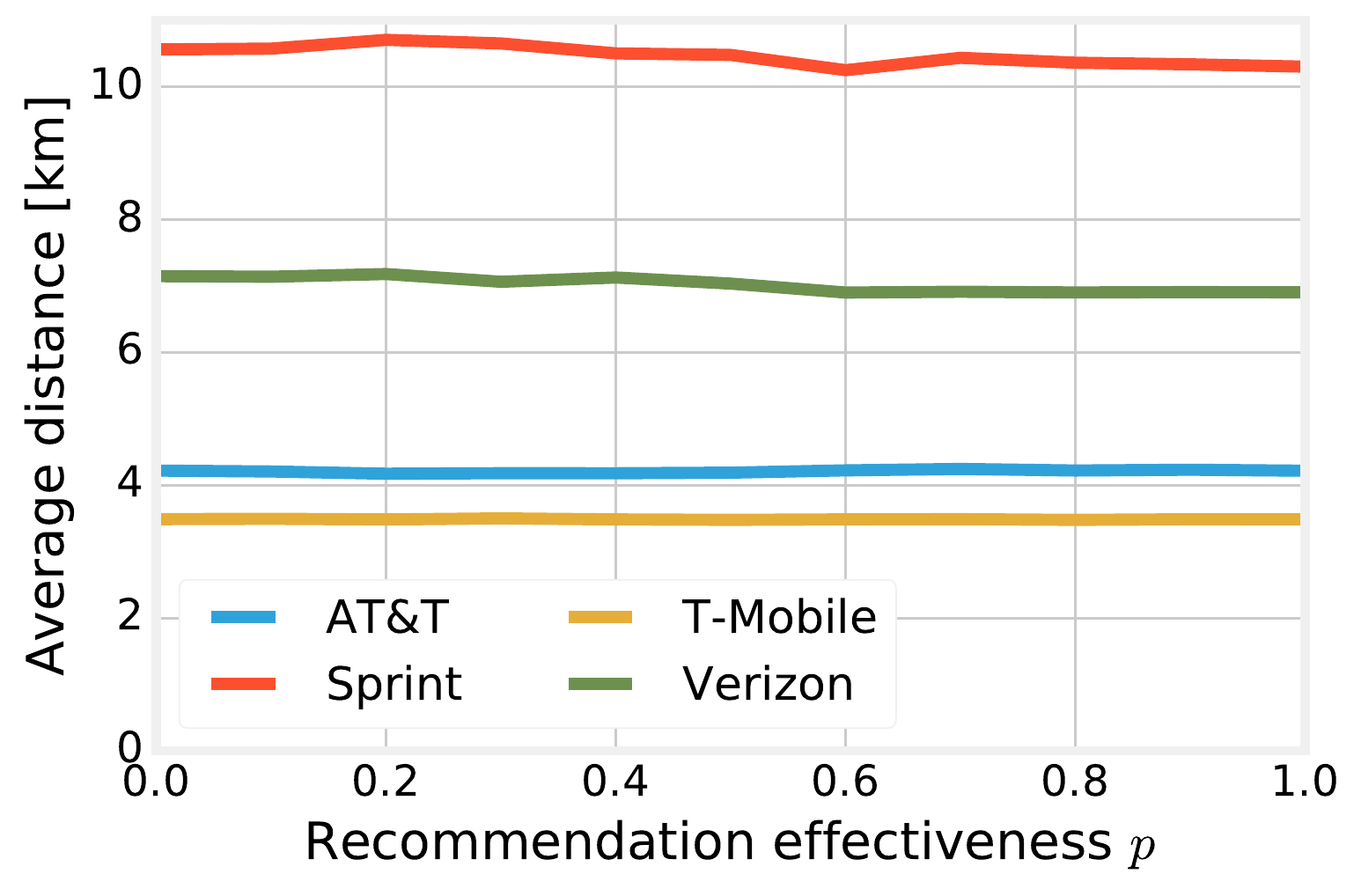}
} 
\caption{
Recommendation system: distance travelled by content items as a function of the effectiveness~$p$ for different operators, when caches are deployed at the rings (a), aggregation pods (b), or core switches (c).
\label{fig:dist-rec}
} 
\end{figure*}

\Fig{dist-rec} shows how the caching architecture and the effectiveness~$p$ of the recommendation system influence the distance travelled by content items for different operators.
{\color{blue}
We can immediately observe that the main factor influencing the distance is the caching architecture (note that plots have different scales), followed by the topology of each operator's core network. Recommendation effectiveness~$p$ has a comparatively minor effect, easier to observe when caches are deployed at aggregation pods (\Fig{dist-rec-pod}).
} 

This seems to contrast with \Fig{rec-pof} and \Fig{rec-cachesize}, that highlight how~$p$ has a significant effect on the price-of-fog and cache size.
The reason is that throughout our numerical results, we keep the target hit ratio fixed, and deploy the minimum amount of cache necessary to achieve it. In other words, we exploit recommendation systems and content locality to reduce the cache size (i.e., the price of the fog) rather than to enhance the benefits thereof (i.e., data travelling shorter distances).
In principle, it would be possible to follow a different strategy and make the distance travelled by content items shorter, at the cost of a higher price-of-fog and/or a larger cache size.

\begin{figure*}[]
\centering
\subfigure[\label{fig:dist-loc-ring}caches at rings]{
\includegraphics[width=.3\textwidth]{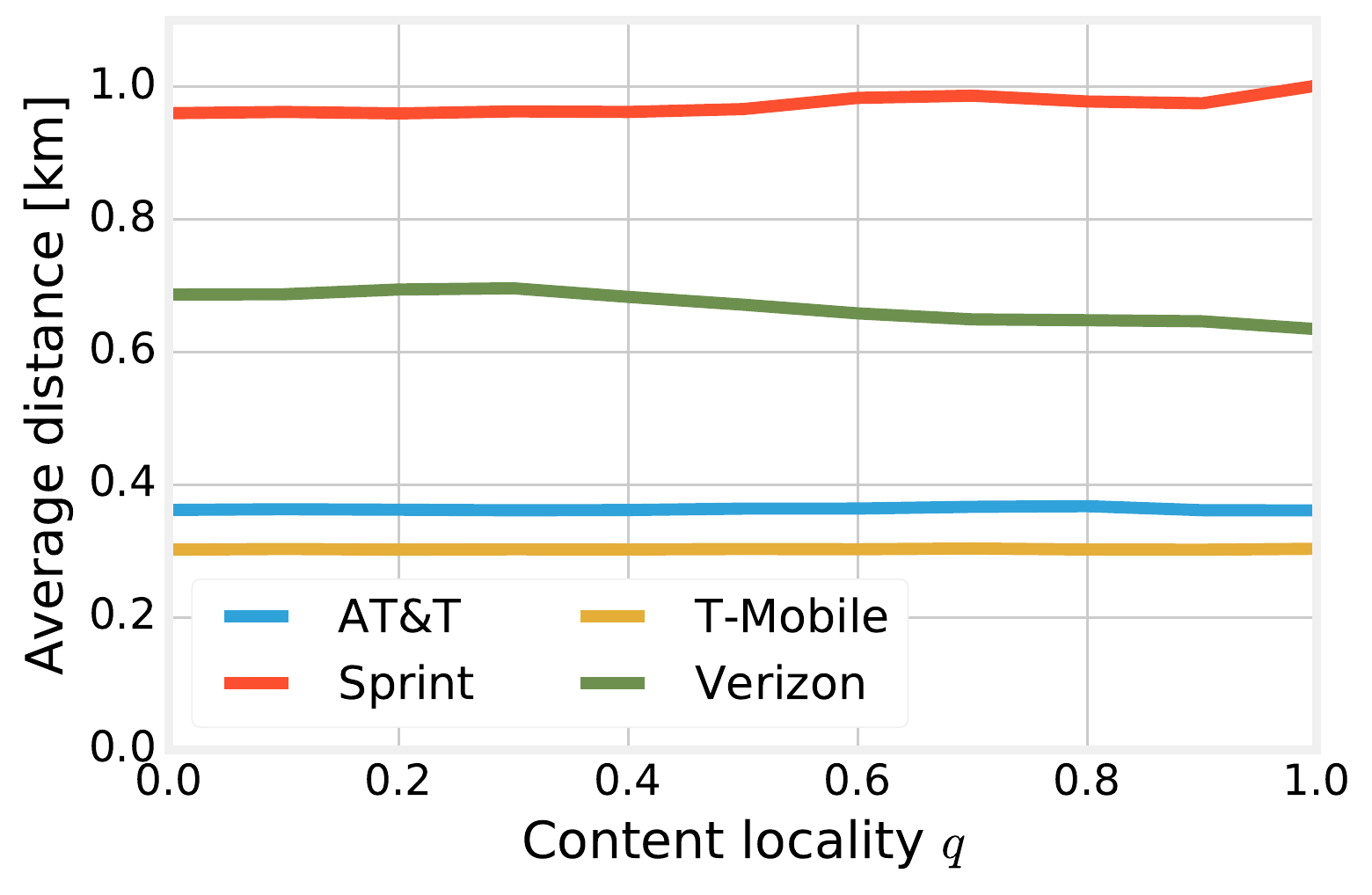}
} 
\subfigure[\label{fig:dist-loc-pod}caches at pods]{
\includegraphics[width=.3\textwidth]{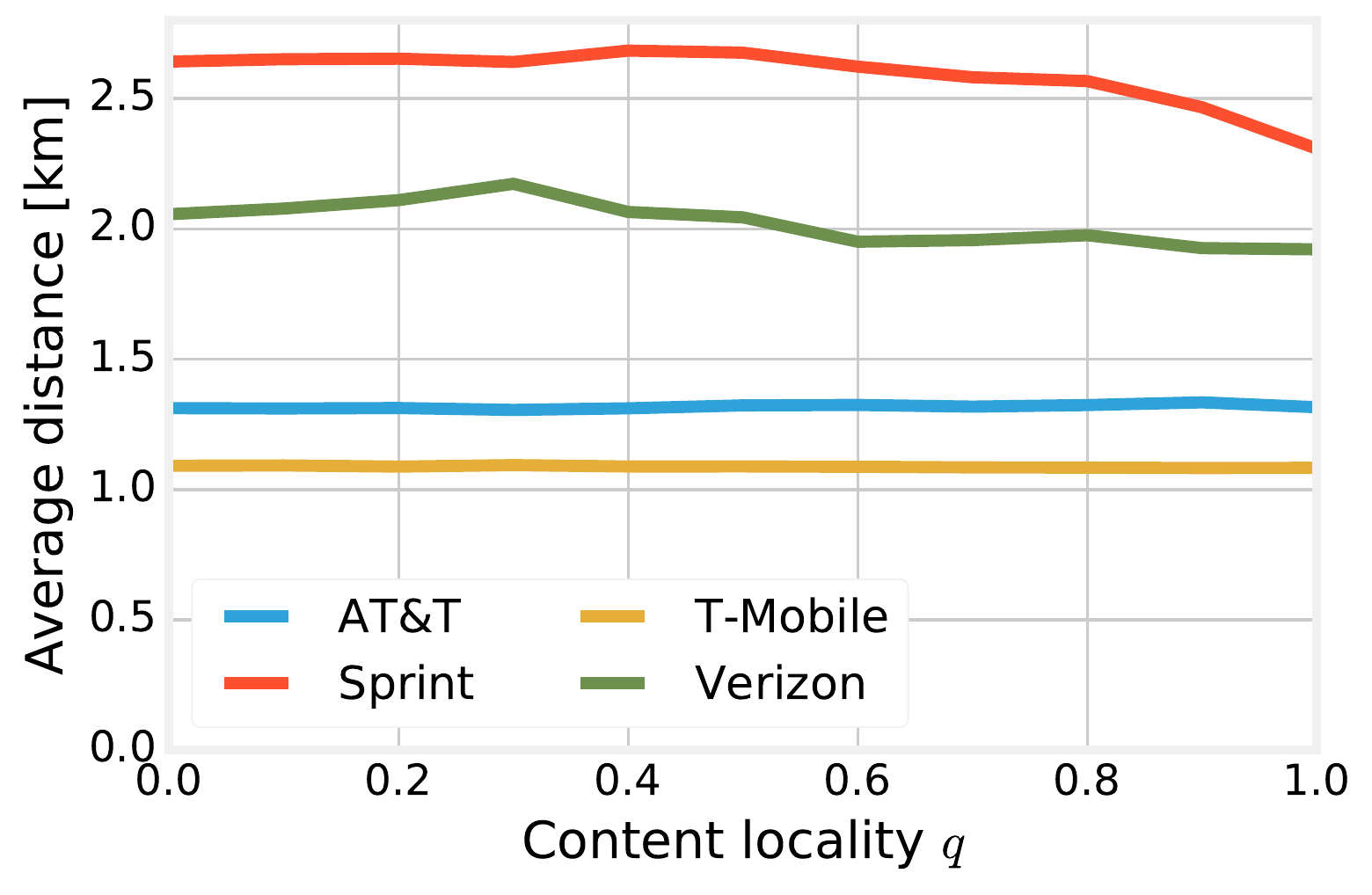}
} 
\subfigure[\label{fig:dist-loc-core}caches at the core]{
\includegraphics[width=.3\textwidth]{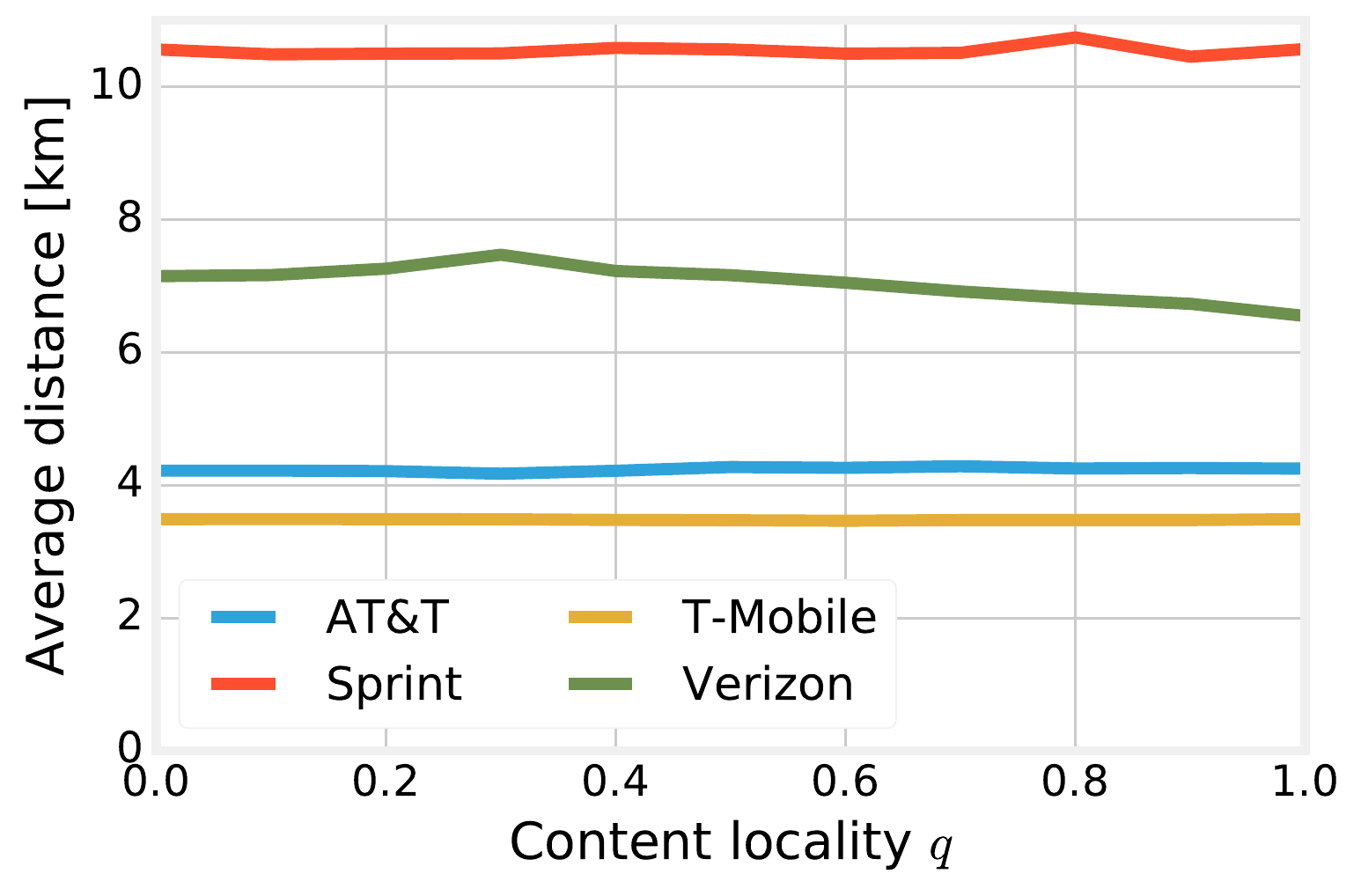}
} 
\caption{
Local content: distance travelled by content items as a function of the locality~$q$ for different operators, when caches are deployed at the rings (a), aggregation pods (b), or core switches (c).
\label{fig:dist-loc}
} 
\end{figure*}

Similar observations can be made about \Fig{dist-loc}, showing the impact of content locality~$q$ on the distance travelled by content items. We again see that the impact is limited when caches are deployed at aggregation pods (\Fig{dist-loc-pod} and almost negligible in all other cases.
{\color{blue}
We consistently find operator topology and caching architecture to be the main factor influencing the distance traveled by data.
} 
As in \Fig{dist-rec}, we can also observe significant differences between operators, stemming from their different network topologies.

{\color{blue}
\paragraph*{Takeaways}

We can make several important observations from our performance evaluation. First, cache sizes are dramatically influenced by the architecture being employed, i.e., at which level caches are placed, and by the individual operators' access and core network topology. Interestingly, the price-of-fog, i.e., the increment in cache size incurred in by moving from a centralized to a more distributed caching architecture, is quite modest, lower than~$1.25$ in all cases.
We further observed that recommendation systems and local contents both influence the price-of-fog, in substantially different ways. Recommendation systems correspond to higher price-of-fog values, as it becomes more likely that multiple copies of the same content get cached. On the other hand, local contents are associated with a lower price-of-fog, a sign that decentralized caching architectures are especially well-suited to this kind of traffic.
} 

\section{Related work}
\label{sec:relwork}

Our paper falls in the general area of caching for mobile cellular networks. The most significant recent trend in this field is {\em fog computing}, also called {\em mobile edge computing}. Compared to traditional cloud computing, the emphasis is to move processing and caching capabilities as close to the access networks (and users) as possible, so as to reduce the load on the core network.

A first body of works deal with the fundamental problem of {\em where} to locate the cached content items, given some degree of knowledge about user demand. For example, the authors of~\cite{commag-air,icn15} exploit concepts from information-centric and content-centric networking to maximize the cache hit ratio, while~\cite{mobaware} leverages mobility information for the same purpose. Other works~\cite{moving} take a more holistic approach, moving both caches and virtual machines around the network as the load changes. All the above build on earlier, more fundamental works; for example, \cite{cac-strat}~uses information density to devise near-optimal caching strategies in ad hoc networks.
{\color{blue}
With respect to these early works, we have the great opportunity to leverage real-world, large-scale deployment and traffic traces, which lends to our conclusions a higher level of realism. At the same time, this faces us with the challenge of obtaining general results, as opposed to results that only apply to the scenario under study.
} 

An especially relevant application of caching is video streaming. As an example,~\cite{multiop-caching,cdn1} account for layered video coding techniques, and address the problem of placing the right layers at the right cache -- with~\cite{multiop-caching} also accounting for cooperation between operators. Other works~\cite{proactive-caching,proseed} aim at {\em foreseeing} the content demand, in order to proactively populate caches~\cite{proactive-caching} or to serve users~\cite{proseed}.

Caching is also a prominent problem in the context of content delivery networks (CDNs). As an example,~\cite{cdn1} studies how to best couple CDNs and mobile networks, considering both technical and economical aspects of the problem. The authors of~\cite{cdn2} study how to optimally deploy CDNs throughout autonomous systems (ASs), and assess how optimization objectives and user requirements influence the performance and cost of caching systems.

Specific to vehicular networks, caching has long been studied in the traditional scenario where no infrastructure support is available~\cite{spawn,mddv}, as well as in infrastructure-powered cases~\cite{noi-tc12}, where a sparse coverage by road-side units (RSUs) exists. More recent works deal with caching in the context of content discovery~\cite{noi-tc12,p2pcache}, content distribution~\cite{content-distrib} and content downloading~\cite{optimal}. A common theme of these works is deploying {\em mobile} caches at the vehicles, and exploiting past and future mobility in order to ensure content availability throughout the network. Our scenario features fixed caches; however, vehicular mobility and traffic patterns do play an important role in defining the caching needs at different parts of the topology.

Closer to our scenario, some recent works~\cite{rsu2015} deal with the problem of caching at RSUs, with the high-level goal of maximizing the hit ratio, given the limited cache size of RSUs and some degree of knowledge about the users' mobility. While RSUs can be seen as analogous to cellular base stations, several important differences exist: cellular networks must serve pedestrian {\em and} vehicular users alike. Furthermore, network operators cannot often rely on user mobility predictions. Finally, we adopt a different performance metric, keeping the hit ratio fixed and studying the cache size we need to obtain it.

{\color{blue}
Finally, a preliminary version of this work has been presented at the IoV-VoI workshop~\cite{iovvoi}. Major improvements with respect to that workshop version include: a more detailed discussion of our scenario and motivation, including the differences between pedestrian and vehicular traffic; an enhanced discussion of cellular network topologies; a new performance evaluation metric, namely, the distance traveled by information within the cellular core network.
} 

\section{Conclusion}
\label{sec:conclusion}

Traffic demand from vehicular users is set to rapidly grow in the next
years, and cellular networks are expected to bear most of the
burden. In this context, we compared different caching
architectures from the viewpoint of the total cache size that operators need to deploy to reach a target hit ratio.

Leveraging a real-world, large-scale, crowd-sourced dataset coming
from the WeFi app, we found that
{\color{blue}
network topology and caching architecture are the main factors influencing the total cache size that mobile operators have to deploy. Furthermore, both highly-localized content and recommendation systems influence such a metric. In particular,
} 
fog computing pairs remarkably well
with highly localized content, such as navigation information for
future self-driving vehicles.
{\color{blue}
On the other hand, traditional recommendation systems do not seem to
combine with mobile-edge caching equally well. Indeed, when the same
content items are popular throughout the whole 
network, more decentralized caching architecture inevitably translate into a higher likelihood of duplicated content.
}

{\color{blue}
\section*{Acknowledgement}
This work has received funding from the 5G-Crosshaul project (H2020-671598), and has been supported by the Israeli-German Cybersecurity Center at the Hebrew University. We would also like to thank the anonymous referees for their constructive suggestions of clarifications in the paper.
} 

\section*{References}
\bibliographystyle{elsarticle-num}
\bibliography{refs}

\end{document}